%% file: v0.tex
\documentclass[prd,onecolumn,superscriptaddress,amssymb,amsfonts,amsmath,nofootinbib,preprintnumbers]{revtex4-2}
\usepackage{graphicx,slashed,xcolor,multirow,bbold,mathtools,sidecap,tikz,bm,ulem,enumitem,booktabs,array}
\usepackage[colorlinks=true,linktoc=page,linkcolor=purple,citecolor=teal,urlcolor=magenta]{hyperref}
\usepackage[compat=1.1.0]{tikz-feynman}
\usepackage{siunitx,adjustbox,comment,soul}
\setstcolor{red}
\definecolor{lime}{HTML}{A6CE39}
\DeclareRobustCommand{\orcidicon}{\hspace{-2.1mm}
\begin{tikzpicture}
\draw[lime,fill=lime] (0,0.0) circle [radius=0.13] node[white] {{\fontfamily{qag}\selectfont \tiny ID}}; \draw[white,fill=white] (-0.0525,0.095) circle [radius=0.007]; 
\end{tikzpicture} \hspace{-3.5mm} }
\foreach \x in {SA,SM} {\expandafter\xdef\csname orcid\x\endcsname{\noexpand\href{https://orcid.org/\csname orcidauthor\x\endcsname} {\noexpand\orcidicon}}}

\let\emph\textit

\begin{document}

\preprint{TTP25-048, P3H-25-102, ICPP-103}

\title{Vacuum Structure of the BNT Model of Neutrino Mass Generation}

\author{Saiyad Ashanujjaman\orcidSA{}}
\email{saiyad.ashanujjaman@kit.edu}
\affiliation{Institut f\"ur Theoretische Teilchenphysik, Karlsruhe Institute of Technology, Engesserstra\ss e 7, D-76128 Karlsruhe, Germany}
\affiliation{Institut f\"ur Astroteilchenphysik, Karlsruhe Institute of Technology, Hermann-von-Helmholtz-Platz 1, D-76344 Eggenstein-Leopoldshafen, Germany}

\author{Siddharth P.~Maharathy\orcidSM{}}
\email{siddharth.prasad.maharathy@cern.ch}
\affiliation{School of Physics and Institute for Collider Particle Physics, University of the Witwatersrand, Johannesburg, Wits 2050, South Africa}
\affiliation{Indian Institute of Science Education and Research Pune, Dr.~Homi Bhabha Road, Pune 411008, India}

\begin{abstract}
We analyze the vacuum structure of the Babu--Nandi--Tavartkiladze (BNT) model of neutrino mass generation, in which the Standard Model is extended by an $SU(2)_L$ scalar quadruplet with hypercharge $Y=3/2$ and a vector-like $SU(2)_L$ triplet fermion with $Y=1$, generating neutrino masses via an effective dimension-seven operator. We delineate the theoretical constraints on the model, requiring the scalar potential to be bounded from below in all field directions, ensuring perturbative unitarity of scattering amplitudes, and demanding that the electroweak vacuum corresponds to the global minimum of the potential. We find that the electroweak vacuum is not generically guaranteed to be the global minimum: several charge-breaking stationary points may coexist with---and potentially lie below---it in potential depth. For the electroweak-like vacuum with vanishing quadruplet expectation value, the condition of global stability reduces to two simple mass inequalities involving the doubly- and triply-charged scalars. In contrast, for the general electroweak vacuum with nonzero doublet and quadruplet expectation values---compatible with neutrino-mass generation---no comparably simple analytic condition emerges, and the stability must in general be assessed for specific choices of scalar couplings. In the special case where the interaction responsible for neutrino-mass generation vanishes, both electroweak configurations coexist, and the bounded-from-below conditions ensure a definite ordering between them. In this limit, the mass inequalities alone are sufficient to guarantee that the general electroweak vacuum is the global minimum. In the physically relevant regime, the results provide practical sufficient criteria and a systematic framework for assessing vacuum stability in the BNT model.
\end{abstract}

\maketitle

\input{sections/intro}
\input{sections/model}
\input{sections/bfb}
\input{sections/unitarity}
\input{sections/globality}
\input{sections/summary}

\acknowledgments
This research was supported by the Deutsche Forschungsgemeinschaft (DFG, German Research Foundation) under grant 396021762 - TRR 257. SPM acknowledges the support of the Research Office of the University of the Witwatersrand. We thank the anonymous referees for careful and thorough reading of the manuscript and for constructive comments that helped improve the presentation, scope, and logical clarity of this work.

\appendix
\input{sections/appA}
\input{sections/appB}
\input{sections/appC}

\bibliography{v0}

\end{document}

%% file: sections/intro.tex
\section{Introduction}
\label{sec:intro}
Among several observational and theoretical lacunae of the Standard Model (SM), of particular concern is the issue of neutrino mass. A well-founded remedy to this shortcoming is offered by the dimension-five Weinberg-operator-induced seesaw mechanism~\cite{Weinberg:1979sa, Ma:1998dn}, wherein lepton-number-violating new physics (NP) beyond the SM is invoked at an a priori unknown scale, so that, on integrating out the heavy NP fields, the SM neutrinos are left with observed sub-eV masses after electroweak (EW) symmetry breaking. In canonical type-I~\cite{Minkowski:1977sc, Yanagida:1979as, Gell-Mann:1979vob, Glashow:1979nm, Mohapatra:1979ia} and type-III~\cite{Foot:1988aq} seesaw models, neutrino masses consistent with data require either an NP scale $\Lambda_{\rm NP}$ close to the grand unification scale for $\mathcal{O}(1)$ Yukawa couplings, or extremely small Yukawa couplings, $\mathcal{O}(10^{-6})$, if $\Lambda_{\rm NP}$ lies around the TeV scale. In the absence of further suppressions (for example, loop suppression or other additional mechanisms), scenarios aiming at collider-accessible NP ($\Lambda_{\rm NP} \sim \text{TeV}$) with sizable Yukawas typically invoke neutrino mass generation via higher-dimensional operators at tree level ~\cite{Bonnet:2009ej, Kanemura:2010bq, Liao:2010ku, Bonnet:2012kz, Cepedello:2017lyo, Anamiati:2018cuq, Ashanujjaman:2020tuv}.

A well-known realization is the model proposed by Babu, Nandi, and Tavartkiladze, hereinafter dubbed the {\it BNT model}~\cite{Babu:2009aq}, where an effective dimension-seven operator induces neutrino masses. The model extends the SM with an $SU(2)_L$ scalar quadruplet of hypercharge $Y=3/2$ and a vectorlike $SU(2)_L$ triplet fermion with $Y=1$. Since its proposal in 2009, this model has been explored in various phenomenological contexts, such as lepton flavor violation~\cite{Liao:2010rx}, interpretations of the 750 GeV diphoton excess~\cite{Ghosh:2016lnu}, collider signatures~\cite{Bambhaniya:2013yca, Babu:2016rcr, Ghosh:2017jbw, Ghosh:2018drw, Pan:2019wwv}, and long-lived particle searches~\cite{Arbelaez:2019cmj}.

In this work, we investigate the vacuum structure of the BNT model. We delineate the theoretical constraints on its scalar potential, requiring it to be bounded from below in all field directions, perturbative unitarity of scattering amplitudes, and the stability of the electroweak (EW) vacuum as the global minimum. In the BNT framework, neutrino masses arise from the dimension-seven operator generated by the $\lambda_{5}\,\Delta^\dagger \Phi^{3}$ interaction. Once the Higgs doublet acquires a VEV, this term induces a tadpole for the neutral component of the scalar quadruplet, implying that $\lambda_{5}\neq 0$ necessarily leads to a nonvanishing quadruplet VEV. As a consequence, only the general electroweak vacuum in which both the doublet and quadruplet develop VEVs is compatible with neutrino-mass generation. The electroweak-like configuration with a vanishing quadruplet VEV exists only for $\lambda_{5}=0$ and cannot generate neutrino masses. This observation directly motivates the detailed vacuum-stability analysis presented here and, in particular, the conditions under which the physically relevant electroweak vacuum constitutes the global minimum of the scalar potential.

While we find that a fully analytic characterization of the global minimum for the general electroweak vacuum is not attainable, the present analysis provides a systematic treatment of the problem. In particular, we classify all stationary configurations of the scalar potential, derive explicit expressions for the potential differences between coexisting extrema, and identify where simple analytic control is possible and where it it is no longer available. Even in the absence of compact analytic conditions for the physically relevant vacuum, the results obtained here reduce the stability problem to a well-defined set of expressions that can be evaluated for a given choice of parameters. This provides a structured basis for numerical analyses and vacuum stability studies in phenomenological applications of the model.

The rest of the paper is organized as follows. In section~\ref{sec:model}, we briefly present the scalar sector of the model. The conditions for the potential to be bounded from below and for perturbative unitarity of scattering amplitudes are discussed in section~\ref{sec:theory}. In section~\ref{sec:globality} we present our analysis of the vacuum structure. Finally, we summarize our findings in section~\ref{sec:summary}.

%% file: sections/model.tex
\section{The BNT Model}
\label{sec:model}
The extended scalar sector of the BNT model is composed of a SM Higgs doublet $\Phi$ with $Y=1/2$ and a Higgs quadruplet $\Delta$ with $Y=3/2$:
\begin{align}
& \Phi = \begin{pmatrix} \phi^+ \\ \phi^0 \end{pmatrix} \quad {\rm and} \quad \Delta = \begin{pmatrix} \delta^{+++} \\ \delta^{++} \\ \delta^+ \\ \delta^0 \end{pmatrix}.
\end{align}
The kinetic part of the Higgs sector Lagrangian is given by
\begin{align}
& \mathcal{L}_{\rm kin} = \left| D_\mu \Phi \right|^2 + \left| D_\mu \Delta \right|^2,
\end{align}
with $D_\mu \Phi = \partial_\mu \Phi - ig \tau^k W_\mu^k \Phi - i\frac{g^\prime}{2} B_\mu \Phi$ and $D_\mu \Delta = \partial_\mu \Delta - ig T^k W_\mu^k \Delta - i\frac{3g^\prime}{2} B_\mu \Delta$, where $W_\mu^k$ and $B_\mu$ are the $SU(2)_L$ and $U(1)_Y$ gauge fields, and $g$ and $g^\prime$ are the corresponding couplings; $\tau^k$ and $T^k$ are, respectively, the generators of 2 and 4 representation of $SU(2)$, where $\tau^k=\sigma^k/2$ with $\sigma^k$ being the Pauli matrices.\footnote{The generators of 4 represenation of $SU(2)$ are not uniquely determined, however, they must satisfy the commutation relations: $[T^k,T^l]=i\epsilon_{klm}T^m$. In a basis with diagonal third generator (often referred as {\it spherical basis}), the corresponding generators are given by
\begin{align}
T^1 &= \frac{1}{2} \left( \begin{array}{cccc}
0 & \sqrt{3} & 0 & 0 \\ \sqrt{3} & 0 & 2 & 0 \\ 0 & 2 & 0 & \sqrt{3} \\ 0 & 0 & \sqrt{3} & 0
\end{array} \right), \quad
T^2 = \frac{i}{2} \left( \begin{array}{cccc}
0 & -\sqrt{3} & 0 & 0 \\ \sqrt{3} & 0 & -2 & 0 \\ 0 & 2 & 0 & -\sqrt{3} \\ 0 & 0 & \sqrt{3} & 0
\end{array} \right), \quad
T^3 = \frac{1}{2} \left( \begin{array}{cccc}
3 & 0 & 0 & 0 \\ 0 & 1 & 0 & 0 \\ 0 & 0 & -1 & 0 \\ 0 & 0 & 0 & -3
\end{array} \right).
\end{align}
}

The most general renormalizable and gauge-invariant scalar potential can be written as\footnote{One can also write $\Phi$ and $\Delta$ as symmetric rank-1 and rank-3 tensors, respectively, as follows:
\begin{align*}
& \Phi^1 = \phi^+,~ \Phi^0 = \phi^0,~ \Phi^*_i = (\Phi^i)^*,~ \Delta^{111} = \Delta^{+++},~ \Delta^{112} = \frac{\Delta^{++}}{\sqrt{3}},~ \Delta^{122} = \frac{\Delta^+}{\sqrt{3}},~ \Delta^{222} = \Delta^0,~ \Delta^*_{ijk} = (\Delta^{ijk})^*.
\end{align*}
In this tensor notation, various terms in Eq.~\eqref{eq:pot} read as
\begin{align*}
& \Phi^\dagger \Phi = \Phi_i^* \Phi^i, \quad \Delta^\dagger \Delta = \Delta_{ijk}^* \Delta^{ijk}, \quad 4 (\Delta^\dagger T^a \Delta)^2 = 18 \Delta_{ijk}^* \Delta^{ijp} \Delta_{mnp}^* \Delta^{mnk} - 9 (\Delta_{ijk}^* \Delta^{ijk})^2,
\\
& 4 (\Phi^\dagger \tau^a \Phi)(\Delta^\dagger T^a \Delta) = 6 \Phi_i^* \Delta^{ijk} \Delta_{jkl}^* \Phi^l - 3 \Phi_i^* \Phi^i \Delta_{ijk}^* \Delta^{ijk}, \quad \Delta^\dagger \Phi^3 = \Delta_{ijk}^* \Phi^i \Phi^j \Phi^k.
\end{align*}
}
\begin{align}
V(\Phi,\Delta) =& -\mu_\Phi^2 \Phi^\dagger \Phi + \mu_\Delta^2 \Delta^\dagger \Delta + \lambda_1 (\Phi^\dagger \Phi)^2 + \lambda_2 (\Delta^\dagger \Delta)^2 + \tilde\lambda_2 (\Delta^\dagger T^k \Delta)^2 \nonumber
\\
& + \lambda_3 (\Phi^\dagger \Phi)(\Delta^\dagger \Delta) + \lambda_4 (\Phi^\dagger \tau^k \Phi)(\Delta^\dagger T^k \Delta) + (\lambda_5 \Delta^\dagger \Phi^3 + {\rm h.c.}),
\label{eq:pot}
\end{align}
where $\mu_\Phi^2$ and $\mu_\Delta^2$ are the mass-squared parameters, and $\lambda_i$ $(i=1,...,5)$ and $\tilde\lambda_2$ are the independent dimensionless couplings.\footnote{We note in passing that the $\tilde\lambda_{2}$ term was absent in the original formulation of the model~\cite{Babu:2009aq} and consequently also omitted in several subsequent  studies~\cite{Bambhaniya:2013yca, Ghosh:2016lnu, Ghosh:2017jbw, Ghosh:2018drw, Pan:2019wwv}. This omission was, however, first pointed out in Ref.~\cite{Liao:2010rx}.} Hermiticity of the potential allows us to assume that all couplings, except for $\lambda_5$, are real. Though $\lambda_5$ can pick up a {\it would-be} CP phase, this phase is unphysical and can always be absorbed through appropriate field redefinitions:
\begin{align}
& \lambda_5 \to \lambda_5 e^{i\omega}, \quad \Phi \to \Phi e^{i\omega^\prime} \quad {\rm and} \quad \Delta \to \Delta e^{i(\omega + 3\omega^\prime)}.
\end{align}
Therefore, without loss of generality, we also assume $\lambda_5$ to be real. Consequently, the potential $V(\Phi,\Delta)$ in Eq.~\eqref{eq:pot} is CP conserving. The neutral components of $\Phi$ and $\Delta$ can be parametrised as 
\begin{align}
& \phi^0 = \frac{v_\Phi + \phi^0_R + i \phi^0_I}{\sqrt{2}} \quad {\rm and} \quad \delta^0 = \frac{v_\Delta + \delta^0_R + i\delta^0_I}{\sqrt{2}},
\end{align}
where $v/\sqrt{2}$ and $v_\Delta/\sqrt{2}$, respectively, are the vacuum expectation values (VEVs) acquired by $\phi^0$ and $\delta^0$, respectively, after EW symmetry breaking, such that
\begin{align}
\sqrt{v_\Phi^2 + 3v_\Delta^2} = 246\,{\rm GeV}. \label{eq:EWVEV}
\end{align}
Minimizing the potential $V(\Phi,\Delta)$, we obtain
\begin{align}
\label{eq:tadpole1}
& \mu_\Phi^2 = \lambda_1 v_\Phi^2 + \frac{\lambda_3}{2} v_\Delta^2 + \frac{3}{8} \lambda_4 v_\Delta^2 + \frac{3}{2} \lambda_5 v_\Phi v_\Delta,
\\
\label{eq:tadpole2}
& \mu_\Delta^2 = -(\lambda_2 +\frac{9}{4} \tilde{\lambda}_2) v_\Delta^2 - \frac{\lambda_3}{2} v_\Phi^2 - \frac{3}{8} \lambda_4 v_\Phi^2 - \frac{\lambda_5}{2}  \frac{v_\Phi^3}{v_\Delta}.
\end{align}
It is worth emphasizing a structural feature of the scalar potential: the quartic interaction $\lambda_{5} \Delta^{\dagger}\Phi^{3}$ induces a linear term in the neutral quadruplet field once the doublet acquires a VEV. Consequently, unless $\lambda_{5}=0$, the minimization condition in Eq.~\eqref{eq:tadpole2} necessarily enforces a nonvanishing quadruplet VEV $v_\Delta$. In other words, for generic parameter choices with $\lambda_{5}\neq 0$---precisely the regime relevant for neutrino-mass generation (see the discussion at the end of this section)---the scalar potential does not admit stationary configurations with $v_{\Delta}=0$.

After EW symmetry breaking, the degrees of freedom with identical electric charges mix, giving rise to several physical Higgs states: 
\begin{enumerate}[label=$(\roman*)$]
\item the neutral states $\phi^0_R$ and $\delta^0_R$ ($\phi^0_I$ and $\delta^0_I$) mix into two CP-even (CP-odd) states $h$ and $H$ ($G$ and $A$), 
\item the singly-charged states $\phi^\pm$ and $\delta^\pm$ mix into two mass states $G^\pm$ and $H^\pm$, 
\item the doubly- and triply-charged states $\delta^{\pm \pm}$ and $\delta^{\pm \pm \pm}$ align with their mass states $H^{\pm \pm}$ and $H^{\pm \pm \pm}$, respectively;
\end{enumerate}
\begin{align*}
& \begin{pmatrix} h \\ H \end{pmatrix} = R_\alpha \begin{pmatrix} \phi^0_R \\ \delta^0_R \end{pmatrix}, ~
\begin{pmatrix} G \\ A \end{pmatrix} = R_\beta \begin{pmatrix} \phi^0_I \\ \delta^0_I \end{pmatrix}, ~
\begin{pmatrix} G^\pm \\ H^\pm \end{pmatrix} = R_\gamma \begin{pmatrix} \phi^\pm \\ \delta^\pm \end{pmatrix} ~{\rm with}~
R_\theta = \begin{pmatrix} \cos\theta & \sin\theta \\ -\sin\theta & \cos\theta \end{pmatrix},
\end{align*}
and
\begin{align*}
R_\alpha^T m_R^2 R_\alpha = {\rm diag} \left(m_h^2,m_H^2\right), ~
R_\beta^T m_I^2 R_\beta = {\rm diag} \left(0,m_A^2\right), ~
R_\gamma^T m_\pm^2 R_\gamma = {\rm diag} \left(0,m_{H^\pm}^2\right),
\end{align*}
where $m_R^2$, $m_I^2$ and $m_\pm^2$, respectively, are the mass matrices for CP-even, CP-odd and singly-charged Higgses:
\begin{align}
& m_R^2 =
\begin{pmatrix}
2 \lambda_1 v_\Phi^2  + \frac{3}{2}\lambda_5 v_\Phi v_\Delta & \lambda_3 v_\Phi v_\Delta  + \frac{3}{4}\lambda_4 v_\Phi v_\Delta + \frac{3}{2}\lambda_5 v_\Phi^2
\\ 
\lambda_3 v_\Phi v_\Delta  + \frac{3}{4}\lambda_4 v_\Phi v_\Delta + \frac{3}{2}\lambda_5 v_\Phi^2 & 2(\lambda_2 + \frac{9}{4}\tilde{\lambda}_2)  v_\Delta^2  - \frac{\lambda_5 v_\Phi^3}{2v_\Delta}
\end{pmatrix}, \label{eq:mR2}
\\
& m_I^2 = \frac{1}{2}\lambda_5 v
\begin{pmatrix}
- 9v_\Delta & 3v_\Phi
\\
3v_\Phi & - \frac{v_\Phi^2}{v_\Delta}
\end{pmatrix},
\\
& m_\pm^2 = (\lambda_4 v_\Delta  + 2 \lambda_5 v_\Phi)
\begin{pmatrix}
- \frac{3}{4}v_\Delta & \frac{\sqrt{3}}{4}v_\Phi
\\
\frac{\sqrt{3}}{4}v_\Phi & -\frac{v_\Phi^2}{4v_\Delta}
\end{pmatrix},
\end{align}
and $\alpha$, $\beta$ and $\gamma$ are the respective mixing angles:
\begin{align}
& \tan 2\alpha = \frac{2\lambda_3 v_\Phi v_\Delta  + \frac{3}{2}\lambda_4 v_\Phi v_\Delta + 3\lambda_5 v_\Phi^2}{2 \lambda_1 v_\Phi^2  -2(\lambda_2 + \frac{9}{4}\tilde{\lambda}_2)  v_\Delta^2 +  \frac{3}{2}\lambda_5 v_\Phi v_\Delta\left(1 + \frac{v_\Phi^2}{3 v_\Delta^2 }\right)},
\\
& \tan 2\beta = \frac{6 v_\Delta}{v_\Phi} \left(1-\frac{9v_\Delta^2}{v_\Phi^2}\right)^{-1},
\\
& \tan 2\gamma = \frac{2\sqrt{3} v_\Delta}{v_\Phi} \left(1-\frac{3v_\Delta^2}{v_\Phi^2}\right)^{-1}.
\end{align}
The mass states $G^0$ and $G^\pm$ are the so-called {\it would-be} Nambu-Goldstone bosons eaten by the longitudinal modes of $Z$ and $W^\pm$, and the rest are massive with $h$ being the 125 GeV Higgs observed at the LHC. The physical Higgs masses are given by\footnote{We note that Eqs.~\eqref{eq:mh_2} and \eqref{eq:mH_2} are obtained from the $(2,2)$ entry of the CP-even mass matrix in Eq.~\eqref{eq:mR2}. These forms involve the combination $\lambda_5/v_\Delta$ and are therefore not suitable for taking the limit $\lambda_5 \to 0$, $v_\Delta \to 0$, where they become singular. In this limit, the physical masses should instead be obtained from Eqs.~\eqref{eq:mh_1} and \eqref{eq:mH_1} or directly from the mass matrix in Eq.~\eqref{eq:mR2}, where no such singularity arises.}
\begin{subequations}
\begin{align}
& m^2_h = \left( 2 \lambda_1 v_\Phi^2  + \frac{3}{2}\lambda_5 v_\Phi v_\Delta \right) + \left(\lambda_3 v_\Phi v_\Delta + \frac{3}{4}\lambda_4 v_\Phi v_\Delta + \frac{3}{2}\lambda_5 v_\Phi^2 \right) \tan\alpha \label{eq:mh_1}
\\
& \quad \quad = 2(\lambda_2 + \frac{9}{4}\tilde{\lambda}_2)  v_\Delta^2 - \frac{\lambda_5 v_\Phi^3}{2v_\Delta} + \left(\lambda_3 v_\Phi v_\Delta + \frac{3}{4}\lambda_4 v_\Phi v_\Delta + \frac{3}{2}\lambda_5 v_\Phi^2 \right) \cot\alpha, \label{eq:mh_2}
\end{align}
\end{subequations}
\begin{subequations}
\begin{align}
& m^2_H = \left( 2 \lambda_1 v_\Phi^2  + \frac{3}{2}\lambda_5 v_\Phi v_\Delta \right) - \left(\lambda_3 v_\Phi v_\Delta + \frac{3}{4}\lambda_4 v_\Phi v_\Delta + \frac{3}{2}\lambda_5 v_\Phi^2 \right) \cot\alpha \label{eq:mH_1}
\\
& \quad \quad = 2(\lambda_2 + \frac{9}{4}\tilde{\lambda}_2)  v_\Delta^2 - \frac{\lambda_5 v_\Phi^3}{2v_\Delta} - \left(\lambda_3 v_\Phi v_\Delta + \frac{3}{4}\lambda_4 v_\Phi v_\Delta + \frac{3}{2}\lambda_5 v_\Phi^2 \right) \tan\alpha, \label{eq:mH_2}
\end{align}
\end{subequations}
\begin{align}
& m^2_A = -\frac{\lambda_5 v_\Phi^3}{2v_\Delta} \left(1+\frac{9v_\Delta^2}{v_\Phi^2}\right),
\\
& m^2_{H^\pm} = -\left( \frac{\lambda_5v_\Phi^3}{2 v_\Delta} + \frac{\lambda_4 v_\Phi^2}{4} \right)\left(1+\frac{3v_\Delta^2}{v_\Phi^2}\right),
\\
& m^2_{H^{\pm\pm}} = - \left( \frac{\lambda_5v_\Phi^3}{2 v_\Delta} + \frac{\lambda_4 v_\Phi^2}{2} + 3\tilde\lambda_2 v_\Delta^2 \right),
\\
& m^2_{H^{\pm\pm\pm}} = - \left( \frac{\lambda_5v_\Phi^3}{2 v_\Delta} + \frac{3\lambda_4 v_\Phi^2}{4} + \frac{9\tilde\lambda_2 v_\Delta^2}{2} \right).
\end{align}
Therefore, we have six free scalar parameters (excluding $m_h$ and $v$) in our model: $v_\Delta$, $\alpha$, $m_H$, $m_A$, $m_{H^\pm}$ and $m_{H^{\pm\pm}}$ or $m_{H^{\pm\pm\pm}}$. As such, the scalar parameters can be traded with these free parameters as follows:
\begin{align}
& \lambda_1 = \frac{1}{2v_\Phi^2}\left[m^2_h\cos^2\alpha + m^2_H\sin^2\alpha +  \frac{3v_\Delta^2}{v_\Phi^2 + 9 v_\Delta^2} m_A^2\right],
\\
& \lambda_2 = \frac{1}{2 v_\Delta^2}\left[m^2_h\sin^2\alpha + m^2_H\cos^2\alpha + \frac{1}{2} \frac{v_\Phi^2}{v_\Phi^2 + 9 v_\Delta^2} m_A^2 - \frac{3v_\Phi^2}{v_\Phi^2 + 3 v_\Delta^2} m^2_{H^\pm} +\frac{3}{2}m^2_{H^{\pm\pm}}\right],
\\
& \tilde\lambda_2 = -\frac{1}{3v_\Delta^2} \left[ m^2_{H^{\pm\pm}} + \frac{v_\Phi^2}{v_\Phi^2 + 9 v_\Delta^2} m^2_A - \frac{2v_\Phi^2}{v_\Phi^2 + 3 v_\Delta^2} m^2_{H^\pm} \right],
\\
& \lambda_3 = \frac{\sin\alpha \cos\alpha}{v_\Phi v_\Delta}(m_h^2-m_H^2) - \frac{3}{v_\Phi^2 + 9 v_\Delta^2}m_A^2 + \frac{6}{v_\Phi^2 + 3 v_\Delta^2}m^2_{H^\pm},
\\
& \lambda_4 = \frac{4}{v_\Phi^2 + 9 v_\Delta^2} m^2_A - \frac{4}{v_\Phi^2 + 3 v_\Delta^2} m^2_{H^\pm},
\\
& \lambda_5 = -\frac{2v_\Delta}{v_\Phi} \frac{1}{v_\Phi^2 + 9 v_\Delta^2} m^2_A.
\end{align}
The remaining two scalar parameters $\mu_\Phi^2$ and $\mu_\Delta^2$ are then related to the free parameters through the tadpole conditions in Eq.~\eqref{eq:tadpole1} and Eq.~\eqref{eq:tadpole2}.\\

\noindent Both $\phi^0$ and $\delta^0$ contribute to the weak gauge boson's masses at tree level:
\begin{align}
& m_W^2 = \frac{1}{4} g^2 (v_\Phi^2 + 3v_\Delta^2) \quad {\rm and} \quad m_Z^2 = \frac{1}{4} (g^2 + g^{'2}) (v_\Phi^2 + 9v_\Delta^2),
\end{align}
that the $\rho = \dfrac{m_W^2}{m_Z^2\cos^2\theta_w}$ parameter takes the form
\begin{align}
& \rho = \frac{v_\Phi^2 + 3v_\Delta^2}{v_\Phi^2 + 9v_\Delta^2}.
\end{align}
The value of the $\rho$ parameter from the EW precision data $\rho = 1.00031(19)$~\cite{ParticleDataGroup:2024cfk} leads to an upper bound of $v_\Delta \sim \mathcal{O}(1)$ GeV, i.e. $v_\Delta \ll v_\Phi$. In this limit,
\begin{align}
& \sin\beta \approx \frac{3 v_\Delta}{v_\Phi},
\\
& \sin\gamma \approx \frac{\sqrt{3} v_\Delta}{v_\Phi},
\end{align}
and
\begin{align}
& m^2_A \approx - \frac{\lambda_5 v_\Phi^3}{2v_\Delta},
\\
& m^2_{H^\pm} \approx - \left( \frac{\lambda_5 v_\Phi^3}{2v_\Delta} + \frac{\lambda_4 v_\Phi^2}{4} \right),
\\
& m^2_{H^{\pm\pm}} \approx - \left( \frac{\lambda_5 v_\Phi^3}{2v_\Delta} + \frac{\lambda_4 v_\Phi^2}{2} \right),
\\
& m^2_{H^{\pm\pm\pm}} \approx - \left( \frac{\lambda_5 v_\Phi^3}{2v_\Delta} + \frac{3\lambda_4 v_\Phi^2}{4} \right),
\end{align}
and their mass-squared differences become
\begin{align}
& m^2_{H^{\pm\pm\pm}} - m^2_{H^{\pm\pm}} \approx m^2_{H^{\pm\pm}} - m^2_{H^\pm} \approx m^2_{H^\pm} - m^2_A \approx -\frac{\lambda_4 v_\Phi^2}{4}.
\end{align}
Further, for small $\alpha$,
\begin{align}
& m^2_h \approx 2 \lambda_1 v_\Phi^2,
\\
& m^2_H \approx m^2_A \approx - \frac{\lambda_5 v_\Phi^3}{2v_\Delta}.
\end{align}

For completeness, we briefly recall how neutrino masses arise in the BNT model. Integrating out the heavy vectorlike fermion triplet and the scalar quadruplet generates the effective dimension-seven operator
\begin{align}
\mathcal{L}_{\rm eff} \supset \frac{c_{ij}}{\Lambda^{3}} \left(\overline{L_i^c}\,\tilde\Phi^\ast\right) \left(\tilde\Phi^\dagger L_j\right) (\Phi^\dagger\Phi) \;+\; {\rm h.c.},
\end{align}
where $L_i$ are the lepton doublets, $\tilde\Phi = i\sigma_2 \Phi^\ast$, and $c_{ij}$ encodes combinations of the underlying Yukawa couplings and heavy masses. After EW symmetry breaking, this operator yields Majorana masses for the light neutrinos:
\begin{align}
(m_\nu)_{ij} \;\sim\; c_{ij} \frac{v_\Phi^{4}}{\Lambda^{3}}\,.
\end{align}

In the renormalizable BNT framework, this operator is generated through the $\lambda_{5}\,\Delta^\dagger\Phi^{3}$ interaction in the scalar potential, together with the Yukawa interactions
\begin{align}
\mathcal{L}_Y \supset Y_i\,\overline{L}_i \Sigma \tilde{\Phi}
+ Y'_i\,\overline{\Sigma}\,\Delta\,L_i + \text{h.c.},
\end{align}
leading to
\begin{align}
(m_\nu)_{ij} = \lambda_{5} \left(Y_i Y_j^\prime + Y_i^\prime Y_j \right)
\frac{v_\Phi^{4}}{M_\Sigma M_\Delta^{2}},
\end{align}
where $M_\Sigma$ is the mass of the vectorlike fermion and $M_\Delta^2 = \mu_\Delta^2 + \left(4\lambda_3 + 3\lambda_4 \right) v_\Phi^2/8+ \left(4\lambda_2 + 9\tilde\lambda_2 \right) 3v_\Delta^2/4$ is the mass-squared of the neutral quadruplet scalar.

Using the minimization conditions in Eqs.~\eqref{eq:tadpole1}–\eqref{eq:tadpole2}, the neutrino masses can equivalently be expressed in terms of the induced quadruplet VEV $v_\Delta$ as
\begin{equation}
(m_\nu)_{ij} \;\propto\;
\left(Y_i Y_j^\prime + Y_i^\prime Y_j \right)
\frac{v_\Phi v_\Delta}{M_\Sigma},
\end{equation}
making explicit that the same $\lambda_{5}$ coupling that induces $v_\Delta$ also controls the overall scale of neutrino masses.

%% file: sections/bfb.tex
\section{Theoretical constraints}
\label{sec:theory}

\subsection{Bounded from below}
\label{sec:bfb} 
A necessary condition for the stability of the vacuum comes from requiring that the potential in Eq.~\eqref{eq:pot} to be bounded from below for all directions in the field space. Obviously, at large field values, the potential in Eq.~\eqref{eq:pot} is dominated by the terms that are quartic in the fields:
\begin{align}
V^{(4)}(\Phi,\Delta) =~& \lambda_1 (\Phi^\dagger \Phi)^2 + \lambda_2 (\Delta^\dagger \Delta)^2 + \tilde\lambda_2 (\Delta^\dagger T^k \Delta)^2 + \lambda_3 (\Phi^\dagger \Phi)(\Delta^\dagger \Delta) + \lambda_4 (\Phi^\dagger \tau^k \Phi)(\Delta^\dagger T^k \Delta) + (\lambda_5 \Delta^\dagger \Phi^3 + {\rm h.c.}).
\end{align}
Defining $x = \left( |\Phi|^2 ~~ |\Delta|^2 ~~ |\Phi||\Delta| \right)^T$ with $|\Phi|^2 = \Phi^\dagger \Phi$, $|\Delta|^2 = \Delta^\dagger \Delta$, and
\begin{align*}
\kappa_1 = \dfrac{(\Delta^\dagger T^k \Delta)^2}{|\Delta|^4}, \quad \kappa_2 = \dfrac{(\Phi^\dagger \tau^k \Phi)(\Delta^\dagger T^k \Delta)}{|\Phi|^2 |\Delta|^2}, \quad \kappa_3 = \dfrac{(\Delta^\dagger \Phi^3 + {\rm h.c.})}{|\Phi|^3 |\Delta|},
\end{align*}
we can rewrite the potential in the following bi-quadratic form
\begin{align}
V(\Phi,\Delta)^{(4)} = x^T \Lambda x,
\end{align}
where
\begin{align}
\Lambda = \left( \begin{array}{ccc}
\lambda_1  & \frac{1}{2}(1-c)(\lambda_3 + \kappa_2\lambda_4) & \frac{1}{2}\kappa_3 \lambda_5
\\
\frac{1}{2}(1-c)(\lambda_3 + \kappa_2\lambda_4) & \lambda_2 +\kappa_1\tilde{\lambda}_2  & 0
\\
\frac{1}{2}\kappa_3\lambda_5 & 0 & c(\lambda_3 + \kappa_2\lambda_4)
\end{array} \right)
\end{align}
with $\kappa_1 \in [0,\frac{9}{4}]$, $\kappa_2 \in [-\frac{3}{4},\frac{3}{4}]$, and $\kappa_3 \in [-2,2]$. The parameter $c$ is an auxiliary variable introduced to parametrize different decompositions of the quartic form. It is not a physical parameter and is taken to lie in the interval $0 \le c \le 1$, ensuring that all relevant copositivity conditions are covered. Now, applying the copositivity conditions~\cite{HADELER198379, CHANG1994113} to $\Lambda$, we find the following conditions:
\begin{align}
&\lambda_1 > 0 \label{eq:lam1}
\\
&\lambda_2 +\kappa_1\tilde{\lambda}_2 > 0 \label{eq:lam2}
\\
&c(\lambda_3 + \kappa_2\lambda_4) > 0
\\
&(1-c)(\lambda_3 + \kappa_2\lambda_4) + 2\sqrt{\lambda_1(\lambda_2+\kappa_1\tilde{\lambda}_2)} > 0
\\
&\kappa_3\lambda_5+2\sqrt{c\lambda_1(\lambda_3 + \kappa_2\lambda_4)} > 0 
\\
&c(1-c)^2(\lambda_3+\kappa_2\lambda_4)^3 + \kappa_3^2\lambda_5^2(\lambda_2 + \kappa_1\tilde{\lambda}_2) - 4c\lambda_1(\lambda_2 + \kappa_1\tilde{\lambda}_2)(\lambda_3+\kappa_2\lambda_4) > 0.
\end{align}

These conditions must be respected for all allowed values of $\kappa_{1,2,3}$ and $c$ for the potential to be bounded from below. Since $\kappa_1$ and $\kappa_2$ individually lie in the ranges $[0,\frac{9}{4}]$ and $[-\frac{3}{4},\frac{3}{4}]$, respectively, one might expect that the allowed region in the $(\kappa_1,\kappa_2)$ plane is the full rectangle spanned by these intervals. This is, however, not the case: $\kappa_1$ and $\kappa_2$ are correlated and must satisfy
\[
|\kappa_2| \leq \frac{1}{2} \sqrt{\kappa_1},
\]
so that the true parameter space is confined by the parabola above.\footnote{The relation between $\kappa_1$ and $\kappa_2$ follows directly from their definitions and the Cauchy--Schwarz inequality. Writing $a^k \equiv \Phi^\dagger \tau^k \Phi$ and $b^k \equiv \Delta^\dagger T^k \Delta$, one has $\kappa_1 = (b^k b^k)/|\Delta|^4$ and $\kappa_2 = (a^k b^k)/(|\Phi|^2 |\Delta|^2)$. Applying $|a^k b^k| \le \sqrt{a^k a^k}\,\sqrt{b^k b^k}$, one obtains $|\kappa_2| \le \frac{1}{2}\sqrt{\kappa_1}$.} For $\kappa_3$, the only restriction is the global bound 
\[
|\kappa_3| \;\leq\; 2,
\] 
and it can vary essentially independently of $\kappa_1,\kappa_2$. Hence, the full three-dimensional parameter space is obtained by extruding the $(\kappa_1,\kappa_2)$ region along the $\kappa_3$ axis between $-2$ and $+2$, forming a ``parabolic wedge prism" in $(\kappa_1,\kappa_2,\kappa_3)$ space.

%% file: sections/unitarity.tex
\subsection{Perturbative unitarity}
\label{sec:unitarity}
Requiring perturbative unitarity in scattering processes constrains the size of the interactions. The partial-wave decomposition of the scattering amplitude $\mathcal{M}_{i\to f}$ reads as 
\begin{equation}
 \mathcal{M}_{fi} = iT^{fi} = 16i\pi \sum_j (2j+1) a^{fi}_j(s) P_j(\cos\theta),   
\end{equation}
where $a_j$ denotes the $j$-th partial-wave amplitude, $\theta$ is the polar angle between
the initial $i$ and final $f$ states, and $P_j$ is the Legendre polynomial of degree $j$. In the high energy (massless) limit, the dominant contribution comes from the $j=0$ partial-wave ($s$-wave) at tree-level
\begin{equation}
a_0^{fi} = -\frac{i}{16\pi} \mathcal{M}_{fi}.
\end{equation}

Unitarity of the $S$-matrix requires that $|a_0| \leq 1$, $|{\rm Re}(a_0)| \leq 1/2$, and $0 \leq {\rm Im}(a_0) \leq 1$. In practice, it is sufficient to impose either $|a_0| \leq 1$ or $|{\rm Re}(a_0)| \leq 1/2$, which in turn implies that the eigenvalues $x_i$ of the scattering submatrices satisfy $|x_i| \leq \kappa \pi$, with $\kappa = 16$ or 8 depending on the chosen criterion. Following \cite{Logan:2022uus}, we adopt the former condition.

By virtue of the equivalence theorem~\cite{Pal:1994jk, Horejsi:1995jj}, unphysical scalar states can be used in place of the longitudinal components of the gauge bosons in the high-energy limit. Compared to $2 \to 2$ scattering, $2 \to 3$ partial-wave amplitudes can be neglected, as they scale inversely with the energy. Moreover, amplitudes involving trilinear vertices are generally suppressed due to intermediate propagators. Therefore, we focus on two-body, tree-level scalar-scalar scattering processes dominated by quartic interactions.

The number of all possible $q$-charged $2\to 2$ channels constructed from $n$ neutral, $s$ singly-charged, $d$ doubly-charged, and $t$ triply-charged fields are $q=0:\tfrac{n(n+1)}{2}+s^2+d^2+t^2$, $q=1:ns+sd+dt$, $q=2:nd+st+\tfrac{s(s+1)}{2}$, $q=3:nt+sd$, $q=4:st+\tfrac{d(d+1)}{2}$, $q=5:dt$, $q=6:\tfrac{t(t+1)}{2}$. In the present model, $n=4$, $s=2$, $d=1$ and $t=1$. We present the resulting submatrices $\mathcal{M}_q$ structured in terms of net electric charge $q$ in the initial/final states, with their entries corresponding to the quartic couplings that mediate the scalar-scalar scattering processes in Appendix~\ref{app:scattering}. Now, requiring the moduli of the eigenvalues of the submatrices to be $\leq \kappa \pi$, we obtain
\begin{align*}
& \left|4\lambda_2 - 11\tilde\lambda_2 \right| \leq 2\kappa\pi, \quad \left|4\lambda_2 \pm 9\tilde\lambda_2 \right| \leq 2\kappa\pi, \quad \left|4\lambda_2 + 3\tilde\lambda_2 \right| \leq 2\kappa\pi, \quad \left|4\lambda_3 \pm 3\lambda_4 \right| \leq 4\kappa\pi, \quad \left|4\lambda_3 - 5\lambda_4 \right| \leq 4\kappa\pi,
\\
& \left|16 \lambda_1 + 4 \lambda_3 + 5 \lambda_4 \pm 8 \sqrt{(16 \lambda_1 - 4 \lambda_3 - 5 \lambda_4)^2 + 3072 \lambda_5^2} \right| \leq 8\kappa\pi,
\\
& \left|12\lambda_2 -\tilde \lambda_2 \pm \sqrt{(16 \lambda_2^2 - 40 \lambda_2 \tilde \lambda_2 + 793 {\tilde \lambda_2}^2} \right| \leq 4\kappa\pi,
\\
& \left|4\lambda_1 + 4\lambda_2 + 31 \tilde{\lambda}_2 \pm\sqrt{(4\lambda_1 - 4\lambda_2 - 31 \tilde{\lambda}_2)^2 + 40 \lambda_4^2} \right| \leq 4\kappa\pi,
\\
& \left|8\lambda_1 + 4\lambda_3 + 5 \lambda_4 \pm\sqrt{(8\lambda_1 - 4\lambda_3 - 5 \lambda_4)^2 + 1536 \lambda_5^2} \right| \leq 8\kappa\pi,
\\
& \left|12\lambda_1 + 20 \lambda_2 + 15\tilde{\lambda}_2 \pm \sqrt{(12\lambda_1 -20 \lambda_2 - 15\tilde{\lambda}_2)^2 + 128 \lambda_3^2} \right| \leq 4\kappa\pi.
\end{align*}
These conditions ensure that perturbative unitarity is respected in all $2\to 2$ scalar scattering processes and put non-trivial constraints on the parameter space.

%% file: sections/globality.tex
\subsection{Stability against charge-breaking minima}
\label{sec:globality}
In Sec.~\ref{sec:model}, while obtaining the scalar spectrum of the model, we have implicitly assumed that the EW symmetry is spontaneously broken at an electrically neutral configuration in field space, and that this configuration corresponds to the global minimum of the potential. Although the tadpole conditions in Eq.~\eqref{eq:tadpole1} and Eq.~\eqref{eq:tadpole2} ensure that the chosen vacuum represents an extremum of the potential in Eq.~\eqref{eq:pot}, it remains necessary to verify that this extremum is indeed stable---i.e., that it corresponds to a true minimum rather than a saddle point or local maximum.

The scalar potential in Eq.~\eqref{eq:pot} can, in general, admit multiple stationary solutions---both neutral and charge-breaking. In what follows, we identify all stationary points and determine sufficient conditions under which the phenomenologically viable EW vacuum constitutes the global minimum by comparing the potential depths of all extrema.\footnote{While a metastable vacuum with a lifetime longer than the age of the Universe could, in principle, be acceptable, assessing this would require a dedicated tunnelling analysis, which is beyond the scope of this work. We therefore restrict to the stronger requirement that the EW vacuum lies at the global minimum of the potential {\it within the class of stationary configurations with $\langle\phi^0\rangle \neq 0$}. A comparison with electroweak-symmetric extrema with $\langle\phi^0\rangle = 0$ can be imposed separately when absolute stability of the full potential is required.} In particular, we compute the potential differences between the desired electroweak vacuum and the remaining extrema: $\Delta V(\rm{CB};\rm{EW}) \equiv V_{\rm CB} - V_{\rm EW}$. With this definition, $\Delta V(\rm{CB};\rm{EW}) > 0$ implies that the electroweak vacuum is deeper and thus stable against tunnelling into the configuration $X$, whereas $\Delta V(\rm{CB};\rm{EW}) < 0$ indicates the presence of a lower, potentially charge-breaking minimum.

Regarding the gauge choice, three real scalar degrees of freedom can always be absorbed through an appropriate gauge fixing. We work in the unitary gauge, where the scalar doublet reduces to a single real neutral component, so its VEV can, without loss of generality, be taken as real and electrically neutral. The only phase-sensitive term in the CP-conserving potential is the quartic interaction $\lambda_5(\Delta^\dagger \Phi^3 + \text{h.c.})$, while all quadratic and quartic terms depend only on the moduli of the fields. Consequently, the potential depends on phases only through a single linear combination. Minimization with respect to the phases therefore enforces this combination to take values $0$ or $\pi$, both for neutral and charge-breaking configurations. As a result, for all stationary configurations---neutral and charge-breaking---there exists a gauge- and symmetry-equivalent representative that can be represented by real VEVs without loss of generality.

The scalar potential of the BNT model admits several stationary configurations, summarized in Table~\ref{Tab:vacua}. Among these, three correspond to charge-conserving extrema---denoted $N1$, $N2$, and $N3$---referred to as normal minima, where only neutral field components acquire nonzero VEVs. In addition, the potential features fourteen charge-breaking stationary points (denoted $CB1$ to $CB14$), characterized by nonvanishing VEVs of charged fields. Such charge-breaking configurations would spontaneously break electromagnetic gauge invariance and generate a photon mass, rendering them phenomenologically unacceptable.

\begin{table}[htb!]
\begin{center}
\begin{tabular}{|c|c|c|c|c|c|}
\toprule
Vacuum & $\langle \sqrt{2} \phi^0\rangle$ & $\langle \sqrt{2} \delta^0\rangle$ & $\langle \sqrt{2} \delta^+\rangle$ & $\langle \sqrt{2} \delta^{++}\rangle$ & $\langle \sqrt{2} \delta^{+++}\rangle$ \\
\midrule
$N1$ & $v_\Phi$ & $v_\Delta$ & 0 & 0 & 0 \\
\hline
$N2$ & $v_\Phi^\prime$ & 0 & 0 & 0 & 0 \\
\hline
$N3$ & 0 & $v_\Delta^\prime$ & 0 & 0 & 0 \\
\hline
$CB1$ & $v$ & 0 & $v_1$ & 0 & 0 \\
\hline
$CB2$ & $v$ & 0 & 0 & $v_2$ & 0 \\
\hline
$CB3$ & $v$ & 0 & 0 & 0 & $v_3$ \\
\hline
$CB4$ & $v$ & 0 & $v_1$ & $v_2$ & 0 \\
\hline
$CB5$ & $v$ & 0 & $v_1$ & 0 & $v_3$ \\
\hline
$CB6$ & $v$ & 0 & 0 &  $v_2$ & $v_3$ \\
\hline
$CB7$ & $v$ & 0 & $v_1$ &  $v_2$ & $v_3$ \\
\hline
$CB8$ & $v$ & $v_0$ & $v_1$ & 0 & 0 \\
\hline
$CB9$ & $v$ & $v_0$ & 0 & $v_2$ & 0 \\
\hline
$CB10$ & $v$ & $v_0$ & 0 & 0 & $v_3$ \\
\hline
$CB11$ & $v$ & $v_0$ & $v_1$ & $v_2$ & 0 \\
\hline
$CB12$ & $v$ & $v_0$ & $v_1$ & 0 & $v_3$ \\
\hline
$CB13$ & $v$ & $v_0$ & 0 &  $v_2$ & $v_3$ \\
\hline
$CB14$ & $v$ & $v_0$ & $v_1$ &  $v_2$ & $v_3$ \\
\bottomrule
\end{tabular}
\end{center}
\caption{Stationary points of the BNT potential: three charge-conserving and fourteen charge-breaking configurations.}
\label{Tab:vacua}
\end{table}

For clarity, we note that the VEV symbols $v, v_0, v_1, v_2, v_3$ are reused for notational simplicity across the different stationary configurations. These VEVs should not be interpreted as identical across the various stationary points: their numerical values are, in general, distinct for each configuration. The notation is merely a convenient shorthand.

A brief discussion of some stationary configurations is in order. The stationary point \textit{N1} corresponds to the desired EW vacuum, whose structure has already been presented in Sec.~\ref{sec:model}. The configuration \textit{N2} also represents an EW-type vacuum. However, for generic parameters, this is not a true stationary point: the $\lambda_5\,\Delta^\dagger \Phi^3$ interaction induces a tadpole for $\delta^0$ once $\Phi$ acquires a VEV, so the condition $v_\Delta = 0$ can be satisfied only if $\lambda_5 = 0$. Thus, \textit{N2} exists as a genuine extremum only if $\lambda_5 = 0$, and otherwise corresponds to the $v_\Delta \to 0$ boundary of \textit{N1}. The stationary point \textit{N3} corresponds to a vacuum structure in which only $\delta^0$ acquires a VEV. Since the absence of a doublet VEV would imply massless fermions, this extremum is incompatible with observations. For this reason, we do not include any additional stationary configurations with $\langle \phi^0 \rangle = 0$ in Table~\ref{Tab:vacua} and do not consider them further in our analysis. Accordingly, statements about global stability in the following refer to the global minimum {\it within the electroweak-breaking sector}. Before proceeding with the comparison of potential depths, we refer to Appendix~\ref{app:min} for the minimization conditions associated with all stationary configurations.

Now, to calculate the difference in potential depth between two coexisting minima, we follow the bilinear formalism detailed in Refs.~\cite{Ferreira:2004yd, Barroso:2005sm, Kannike:2012pe}; see also Refs.~\cite{Ferreira:2019hfk, Azevedo:2020mjg, Hundi:2023tdq} for recent works. The potential $V$ in Eq.~\eqref{eq:pot} can be decomposed into homogeneous functions of order two, and four in the fields $\Phi$ and $\Delta$: $V = V_2 + V_4$. Consequently, the potential can be written as a quadratic polynomial in the real bilinear vector $X$ constructed from independent field bilinears: $x_1 = |\Phi|^2$, $x_2 = |\delta^0|^2$, $x_3 = |\delta^+|^2$, $x_4 = |\delta^{++}|^2$, $x_5 = |\delta^{+++}|^2$, $x_6 = {\rm Re}(\phi^{0*}\delta^0)$, $x_7 = {\rm Re}(\delta^{0*}\delta^{++})$, $x_8 = {\rm Re}(\delta^{+*}\delta^{+++})$
\begin{align}
V &= \underbrace{M^T X}_{V_2} + \underbrace{\frac{1}{2} X^T \Lambda X}_{V_4},
\end{align}
where $X = (x_1,...,x_8)^T$,
\begin{align}
M =
\begin{pmatrix}
-\mu_\Phi^2 \\ \mu_\Delta^2 \\ \mu_\Delta^2 \\ \mu_\Delta^2 \\ \mu_\Delta^2 \\ 0 \\ 0 \\ 0
\end{pmatrix},
\quad
\Lambda =
\begin{pmatrix}
2\lambda_1 & \lambda_3 + \tfrac{3}{4}\lambda_4 & \lambda_3 + \tfrac{1}{4}\lambda_4 & \lambda_3 - \tfrac{1}{4}\lambda_4 & \lambda_3 - \tfrac{3}{4}\lambda_4 & 2\lambda_5 & 0 & 0 \\[4pt]
\lambda_3 + \tfrac{3}{4}\lambda_4 & 2\lambda_2 + \tfrac{9}{2}\tilde{\lambda}_2 & 2\lambda_2 + \tfrac{9}{2}\tilde{\lambda}_2 & 2\lambda_2 - \tfrac{3}{2}\tilde{\lambda}_2 & 2\lambda_2 - \tfrac{9}{2}\tilde{\lambda}_2 & 0 & 0 & 0 \\[4pt]
\lambda_3 + \tfrac{1}{4}\lambda_4 & 2\lambda_2 + \tfrac{9}{2}\tilde{\lambda}_2 & 2\lambda_2 + \tfrac{1}{2}\tilde{\lambda}_2 & 2\lambda_2 + \tfrac{7}{2}\tilde{\lambda}_2 & 2\lambda_2 - \tfrac{3}{2}\tilde{\lambda}_2 & 0 & 4\sqrt{3}\,\tilde{\lambda}_2 & 0 \\[4pt]
\lambda_3 - \tfrac{1}{4}\lambda_4 & 2\lambda_2 - \tfrac{3}{2}\tilde{\lambda}_2 & 2\lambda_2 + \tfrac{7}{2}\tilde{\lambda}_2 & 2\lambda_2 + \tfrac{1}{2}\tilde{\lambda}_2 & 2\lambda_2 + \tfrac{9}{2}\tilde{\lambda}_2 & 0 & 0 & 4\sqrt{3}\,\tilde{\lambda}_2 \\[4pt]
\lambda_3 - \tfrac{3}{4}\lambda_4 & 2\lambda_2 - \tfrac{9}{2}\tilde{\lambda}_2 & 2\lambda_2 - \tfrac{3}{2}\tilde{\lambda}_2 & 2\lambda_2 + \tfrac{9}{2}\tilde{\lambda}_2 & 2\lambda_2 + \tfrac{9}{2}\tilde{\lambda}_2 & 0 & 0 & 0 \\[4pt]
2\lambda_5 & 0 & 0 & 0 & 0 & 0 & 0 & 0 \\[4pt]
0 & 0 & 4\sqrt{3}\,\tilde{\lambda}_2 & 0 & 0 & 0 & 0 & 6\tilde{\lambda}_2 \\[4pt]
0 & 0 & 0 & 4\sqrt{3}\,\tilde{\lambda}_2 & 0 & 0 & 6\tilde{\lambda}_2 & 0
\end{pmatrix}.
\end{align}

At any stationary point (denoted by \textit{SP}), the potential satisfies the minimization conditions $\partial V/\partial \phi_i = 0$, where $\phi_i$ represents the real scalar components of the model. Multiplying these equations by $\phi_i$ and summing over all components yields
\begin{align} 
& \sum_i \phi_i\frac{\partial V}{\partial \phi_i}\bigg|_{SP} = 0 
\quad 
\Rightarrow 2(V_2)_{SP} + 4(V_4)_{SP} = 0,
\end{align}
in accordance with Euler’s theorem for homogeneous functions. Therefore, the value of the potential at a stationary point can be expressed as
\begin{align}
V_{SP} &= \frac{1}{2} (V_2)_{SP} = - (V_4)_{SP}.
\end{align}
Next, we define the gradient of $V$ with respect to $X$ as 
\begin{align}
V' = \dfrac{\partial V}{\partial X^T} =  M + \Lambda X.
\end{align}
The product of $V'$ evaluated at one stationary point ($SP1$) with $X$ evaluated at another ($SP2$) then satisfies
\begin{align}
&X_{SP1}^T V^\prime_{SP2} = 2V_{SP1} + X_{SP1}^T \Lambda X_{SP2},
\\
&X_{SP2}^T V^\prime_{SP1} = 2V_{SP2} + X_{SP2}^T \Lambda X_{SP1}
\end{align}
Subtracting these two relations and noting that the matrix $\Lambda$ is symmetric, we obtain the general expression for the difference in potential depth between two stationary points:
\begin{equation}
V_{SP2} - V_{SP1} = \frac{1}{2} \left(X_{SP2}^T V^\prime_{SP1} - X_{SP1}^T V^\prime_{SP2}\right).
\label{eq:diff-pot}
\end{equation}

In the following, we analyze the stability of the EW stationary points---i.e. the desired electroweak vacua---against the charge-breaking extrema. Assuming that $N1$ or $N2$ coexists with one or more $CB$ stationary points, we evaluate the corresponding potential differences using Eq.~\eqref{eq:diff-pot}.

We emphasize that the comparison involving $N2$ is restricted to the special case $\lambda_5 = 0$, where $N2$ is a genuine stationary point. In this limit, the absence of the neutral tadpole does not preclude charge-breaking stationary configurations with nonvanishing quadruplet VEVs, as follows from the corresponding minimization conditions.

\paragraph{Stability of $N2$ against charge-breaking extrema.---} The potential differences between $N2$ and the charge-breaking stationary points are provided in Appendix~\ref{app:potdiff}. If $N2$ is a true minimum, all mass-squared eigenvalues are positive. Moreover, as argued above, \textit{N2} exists as a stationry point only if $\lambda_5 = 0$. Consequently, several of the potential differences are manifestly positive, namely $V_{CB2} - V_{N2} > 0$, $V_{CB3} - V_{N2} > 0$, and $V_{CB6} - V_{N2} > 0$. Thus, $N2$ is necessarily stable against charge-breaking extrema of types $CB2$, $CB3$ and $CB6$. For the remaining configurations, however, $N2$ is not guaranteed to be the global minimum, as some charge-breaking stationary points may coexist with---and potentially lie below---it in potential depth. Ensuring that $N2$ is stable against all charge-breaking extrema then requires the mass inequalities
\begin{align}
2m_{H^{\pm\pm}}^2 - m_{H^{\pm\pm\pm}}^2 > 0, \quad 3m_{H^{\pm\pm}}^2 - 2m_{H^{\pm\pm\pm}}^2 > 0. \label{eq:N2vsCB}
\end{align}

\paragraph{Stability of $N1$ vs $N2$.---} When $\lambda_5 = 0$, both $N1$ and $N2$ are genuine stationary points of the potential. In this limit, the minimization conditions for $N1$ remain well defined and admit solutions with $v_\Delta \neq 0$, so that both configurations can coexist and be compared. The difference in potential depth is then given by
\begin{align*}
V_{N2} - V_{N1} =& \frac{1}{4} \left\{ \lambda_1 v_\Phi^2 (v_\Phi^{\prime 2} -  v_\Phi^2) + v_\Delta^4 \left(\lambda_2 + \frac{9}{4} \tilde\lambda_2 \right) - \lambda_5 v_\Phi^3 v_\Delta \right\},
\end{align*}
Since $N2$ exists only for $\lambda_5 = 0$, the last term vanishes. Using the electroweak symmetry–breaking conditions $\sqrt{v_\Phi^2 + 3v_\Delta^2} \approx 246~{\rm GeV}$ (Eq.~\eqref{eq:EWVEV}) and $v_\Phi^\prime \approx 246~{\rm GeV}$, one has $v_\Phi^2 + 3v_\Delta^2 = v_\Phi^{\prime 2}$, so that the condition for $N1$
to be deeper than $N2$ simplifies to
\begin{align}
3\lambda_1 v_\Phi^2 + v_\Delta^2 \left(\lambda_2 + \frac{9}{4} \tilde\lambda_2 \right) > 0. \label{eq:N1vsN2}
\end{align}
This condition is automatically satisfied once the bounded-from-below conditions $\lambda_1 > 0$ and $\lambda_2 + \frac{9}{4}\tilde{\lambda}_2 > 0$ (Eqs.~\eqref{eq:lam1}–\eqref{eq:lam2}) are imposed. Hence, $N1$ is necessarily deeper than $N2$.

We stress that Eq.~\eqref{eq:N2vsCB}, together with the automatic ordering $V_{N1}<V_{N2}$ implied by Eq.~\eqref{eq:N1vsN2} and the bounded-from-below conditions, provides a simple analytic criterion only in the special case $\lambda_5 = 0$. For $\lambda_5 \neq 0$, these relations should be viewed as providing analytic control in the $\lambda_5 \to 0$ limit and as a useful organizing principle (as discussed further in the context of $N1$ stability), rather than as strict stability conditions in the full theory.\\

\paragraph{Stability of $N1$ against charge-breaking extrema.---} The potential differences between $N1$ and the various charge-breaking configurations are provided in Appendix~\ref{app:potdiff}. Some of these expressions can be further simplified by using the \emph{additional} minimization conditions that apply to the CB configurations---that is, the CB-specific tadpole equations beyond the $\mu_\Phi^{2}$ and $\mu_\Delta^{2}$ relations already imposed at the $N1$ stationary point. However, even after incorporating these extra conditions, the resulting
expressions do not, in general, reduce to particularly useful analytic forms.

For the configuration $CB7$, the minimization conditions imply $\tilde \lambda_2 = 0$, $\lambda_4 = 0$, and the potential difference takes the following form when expressed in terms of the physical masses evaluated at the stationary point $N1$
\begin{align*}
V_{CB7} - V_{N1} = \frac{3v^2 v_\Delta^2 + \left(v_1^2 + v_2^2 + v_3^2 \right) v_\Phi^2}{4 \left( v_\Phi^2 + 9v_\Delta^2 \right)} m_A^2.
\end{align*}
If $N1$ is a true minimum, all mass-squared eigenvalues are positive. Therefore the expression above is strictly positive: $V_{CB7} - V_{N1} > 0$, and thus $N1$ is necessarily stable against charge-breaking extrema of type $CB7$.

For CB14, the minimization conditions set $\tilde \lambda_2 = 0$, $\lambda_4 = 0$ and $\lambda_5 = 0$, so that the potential difference vanishes identically: $V_{CB14} - V_{N1} = 0$. The origin of this degeneracy is straightforward: in this limit, the scalar potential depends only on the invariants $\Phi^\dagger \Phi$ and $\Delta^\dagger \Delta$, and is therefore insensitive to the orientation of the triplet in $SU(2)$ space. Consequently, the apparent charge-breaking VEVs of the triplet can be rotated away, continuously mapping the would-be $CB14$ configuration into the $N1$ one. In other words, $CB14$ ceases to be a genuinely charge-breaking vacuum in this limit. For generic scalar-potential parameters---i.e. whenever at least one of $\tilde \lambda_2, \lambda_4$ and $\lambda_5$ is nonzero---this degeneracy is lifted, and no $CB14$ stationary point coexists with $N1$.

At this point, it becomes clear that $N1$ is invariably stable against the configurations $CB7$ and $CB14$. For the remaining charge-breaking extrema, while it is evident that $N1$ is not generically guaranteed to be the global minimum---since several charge-breaking stationary points may coexist with, and potentially lie below, it in potential depth---the corresponding expressions involve several competing terms and cannot be reduced to simple positivity conditions on masses or couplings. As a result, no useful analytic criterion can be extracted for most cases. Therefore, for these configurations, the stability of $N1$ must be assessed for each specific choice of scalar couplings.\\

\paragraph{Globality of $N1$.---}A more practical route to establishing the globality of $N1$ exploits the special role of $N2$. In the special case $\lambda_5 = 0$---where $N2$ is a genuine stationary point of the potential---Eq.~\eqref{eq:N2vsCB} ensures that $N2$ is stable against all charge-breaking extrema. In the same limit, the condition that $N1$ lies below $N2$, Eq.~\eqref{eq:N1vsN2} is automatically satisfied once the bounded-from-below conditions are imposed. Consequently, in this restricted case, $N1$ is guaranteed to lie below all stationary configurations considered and thus constitutes the global minimum of the scalar potential.

For $\lambda_5 \neq 0$, the configuration $N2$ is no longer a stationary point of the scalar potential and should instead be regarded as the $v_\Delta \to 0$ boundary of the vacuum $N1$. At the level of the original tree-level potential, which is a smooth polynomial function of the scalar fields and depends smoothly on $\lambda_5$, standard implicit-function arguments imply that a non-degenerate stationary point persists as a smooth branch under sufficiently small deformations of parameters. In particular, when the stationary point $N1$ exists and is non-degenerate, it continuously approaches the configuration $v_\Delta \to 0$ as $\lambda_5 \to 0$, i.e.\ $N1(\lambda_5 \to 0) \to N2$. In this restricted sense, comparisons involving $N2$ provide useful heuristic guidance and analytic control in the $\lambda_5 \to 0$ limit, and serve as a convenient organizing principle for the vacuum-structure analysis.

It is important to stress, however, that continuity alone does not constitute a strict proof of global stability at finite $\lambda_5$. In particular, it does not exclude the possibility of vacuum reordering away from the $\lambda_5 \to 0$ regime, nor the appearance of additional stationary points when degeneracies are encountered. For this reason, the relations involving $N2$ are not used as necessary conditions for global stability in the physically relevant $\lambda_5 \neq 0$ theory. All rigorous statements are therefore phrased in terms of sufficient conditions, and the role of $N2$ is explicitly limited to that of a practical and limiting reference configuration, rather than serving as the basis of a strict global-minimum proof.

Finally, it is worth emphasizing the phenomenological relevance of the two electroweak stationary points. Neutrino masses in the BNT model arise from the dimension-seven operator induced by the $\lambda_{5}\,\Delta^\dagger \Phi^{3}$ interaction, which necessarily requires $\lambda_{5}\neq 0$ and hence a nonvanishing quadruplet VEV. As a consequence, only the general electroweak vacuum $N1$ is compatible with neutrino-mass generation, whereas the electroweak-like configuration $N2$---which exists solely for $\lambda_{5}=0$---cannot generate neutrino masses and is therefore not phenomenologically interesting. Accordingly, in the physically relevant regime $\lambda_5 \neq 0$, the stability of the vacuum must be assessed directly for $N1$, with $N2$ serving only as a useful limiting reference configuration in the $\lambda_5 \to 0$ limit.

%% file: sections/summary.tex
\section{Summary and outlook}
\label{sec:summary}
In this work we have undertaken a systematic study of the vacuum structure of the Babu--Nandi--Tavartkiladze (BNT) model of neutrino mass generation, imposing from the outset the general theoretical requirements that the scalar potential be bounded from below in all field directions and that the quartic couplings satisfy perturbative unitarity constraints obtained from scalar--scalar scattering amplitudes. We, then, classified the full set of stationary configurations admitted by the scalar potential, derived the corresponding minimization conditions, and evaluated the potential differences between coexisting extrema. This provides a complete basis for comparing the relative depths of the various stationary points relevant to electroweak symmetry breaking and for determining the conditions under which the electroweak vacuum constitutes the global minimum within this sector.

A first important result is that the electroweak-like vacuum $N2$, characterized by a vanishing quadruplet VEV, is stable against all charge-breaking extrema whenever the two simple mass inequalities collected in Eq.~\eqref{eq:N2vsCB} are satisfied. For the general electroweak vacuum $N1$, in which both the doublet and quadruplet acquire VEVs, no comparably simple analytic condition emerges: most charge-breaking configurations lead to potential differences containing several competing terms, preventing the extraction of generic positivity constraints on masses and/or couplings. Consequently, the stability of $N1$ must in general be assessed for specific choices of scalar couplings.

In the special case $\lambda_5 = 0$, where both $N1$ and $N2$ are genuine stationary points, the condition that $N1$ lies below $N2$ is automatically satisfied once the bounded-from-below conditions are imposed. In this limit, the mass inequalities in Eq.~\eqref{eq:N2vsCB}, which ensure that $N2$ is deeper than all charge-breaking extrema, are therefore sufficient to guarantee that $N1$ is the global minimum of the scalar potential.

Finally, we emphasize the phenomenological significance of these results. Neutrino masses in the BNT model arise through the dimension-seven operator generated by the $\lambda_{5}\,\Delta^\dagger \Phi^{3}$ interaction, which requires $\lambda_{5}\neq 0$ and hence a nonvanishing quadruplet VEV. This singles out $N1$ as the only electroweak vacuum compatible with neutrino-mass generation, while $N2$---which exists only for $\lambda_{5}=0$---cannot produce neutrino masses and is therefore phenomenologically disfavored. Accordingly, in the physically relevant regime $\lambda_5 \neq 0$, the stability of the vacuum must be assessed directly for $N1$, with $N2$ serving only as a useful limiting reference configuration. The results obtained here thus provide practical sufficient criteria and a systematic framework for assessing the global stability of the electroweak vacuum in the BNT model.

%% file: sections/appA.tex
\section{$2\to 2$ scalar-scalar scattering submatrices}
\label{app:scattering}
\begin{itemize}[leftmargin=*]
\item \underline{$q=6$:} $\left( \frac{1}{\sqrt{2}}\delta^{+++}\delta^{+++} \right)$
\[
\mathcal{M}_6 = 2\left(\lambda_2 + \frac{9}{4}\tilde{\lambda}_2\right)
\]

\item \underline{$q=5$:} $\left( \delta^{+++}\delta^{++} \right)$
\[
\mathcal{M}_5 = 2\left(\lambda_2 + \frac{9}{4}\tilde{\lambda}_2\right)
\]

\item \underline{$q=4$:} $\left( \delta^{+++}\phi^+, \delta^{+++}\delta^+, \frac{1}{\sqrt{2}}\delta^{++}\delta^{++} \right)$
\[
\mathcal{M}_4 = \left( \begin{array}{ccc}
\lambda_3 + \frac{3}{4} \lambda_4 & 0 & 0
\\
0 & 2\lambda_2 - \frac{3}{2} \tilde\lambda_2 & 2\sqrt{6} \tilde\lambda_2
\\
0 & 2\sqrt{6} \tilde\lambda_2 & 2\lambda_2 + \frac{1}{2} \tilde\lambda_2
\end{array} \right)
\]

\item \underline{$q=3$:} $\left( \delta^{++}\phi^+, \delta^{+++}\phi_R^0, \delta^{+++}\phi_I^0 \right)$
\[
\mathcal{M}_3^a = \left( \begin{array}{ccc}
\lambda_3 + \frac{1}{4} \lambda_4 & \frac{\sqrt{3}}{2\sqrt{2}} \lambda_4 & i\frac{\sqrt{3}}{2\sqrt{2}} \lambda_4
\\
\frac{\sqrt{3}}{2\sqrt{2}} \lambda_4 & \lambda_3 - \frac{3}{4} \lambda_4 & 0
\\
i\frac{\sqrt{3}}{2\sqrt{2}} \lambda_4 & 0 & \lambda_3 - \frac{3}{4} \lambda_4
\end{array} \right)
\]

\item \underline{$q=3$:} $\left( \delta^{++}\delta^+, \delta^{+++}\delta_R^0, \delta^{+++}\delta_I^0 \right)$
\[
\mathcal{M}_3^b = \left( \begin{array}{ccc}
2\lambda_2 + \frac{7}{2} \tilde\lambda_2 & \frac{3}{\sqrt{2}} \tilde\lambda_2 & i\frac{3}{\sqrt{2}} \tilde\lambda_2 
\\
 \frac{3}{\sqrt{2}} \tilde\lambda_2 & 2\lambda_2 - \frac{9}{2} \tilde\lambda_2 & 0
\\
-i\frac{3}{\sqrt{2}} \tilde\lambda_2 & 0 & 2\lambda_2 - \frac{9}{2} \tilde\lambda_2
\end{array} \right)
\]

\item \underline{$q=2$:} $\left(\delta^+\phi^+, \delta^{++}\phi^0_R, \delta^{++}\phi^0_I, \delta^{+++}\phi^-, \phi^+\phi^+ \right)$
\[
\mathcal{M}_2^a = \left( \begin{array}{ccccc}
\lambda_3 -\frac{1}{4}\lambda_4 & \frac{1}{\sqrt{2}}\lambda_4 & i\frac{1}{\sqrt{2}}\lambda_4 & 0 & 0 \\
\frac{1}{\sqrt{2}}\lambda_4 & \lambda_3 -\frac{1}{4}\lambda_4 & 0 & \frac{\sqrt{3}}{2\sqrt{2}}\lambda_4 & \sqrt{6}\lambda_5 \\
-i\frac{1}{\sqrt{2}}\lambda_4 & 0 & \lambda_3 -\frac{1}{4}\lambda_4 & i\frac{\sqrt{3}}{2\sqrt{2}}\lambda_4 & i\sqrt{6}\lambda_5 \\
0 & \frac{\sqrt{3}}{2\sqrt{2}}\lambda_4 & -i\frac{\sqrt{3}}{2\sqrt{2}}\lambda_4 & \lambda_3 +\frac{3}{4}\lambda_4 & 6\lambda_5 \\
0 & \sqrt{6}\lambda_5 & -i\sqrt{6}\lambda_5 & 6\lambda_5 & 4\lambda_1
\end{array} \right)
\]

\item \underline{$q=2$:} $\left(\delta^+\delta^+, \delta^{++}\delta^0_R, \delta^{++}\delta^0_I, \delta^{+++}\delta^- \right)$
\[
\mathcal{M}_2^b = \left( \begin{array}{cccc}
4\lambda_2 + \tilde \lambda_2 & 2\sqrt{6} \tilde \lambda_2 & i2\sqrt{6} \tilde \lambda_2 & 0 \\
2\sqrt{6} \tilde \lambda_2 & 2\lambda_2 - \frac{3}{2} \tilde \lambda_2 & 0 & \frac{3}{\sqrt{2}} \tilde \lambda_2 \\
-i2\sqrt{6} \tilde \lambda_2 & 0 & 2\lambda_2 - \frac{3}{2} \tilde \lambda_2 & i\frac{3}{\sqrt{2}} \tilde \lambda_2 \\
0 & \frac{3}{\sqrt{2}} \tilde \lambda_2 & -i\frac{3}{\sqrt{2}} \tilde \lambda_2 & 2\lambda_2 - \frac{3}{2} \tilde \lambda_2 \\
\end{array} \right)
\]

\item \underline{$q=1$:} $\left(  \delta^{+++}\delta^{--}, \delta^{++}\delta^-, \delta^+ \delta^0_R, \delta^+ \delta^0_I, \phi^+ \phi^0_R, \phi^+ \phi^0_I, \delta^{++}\phi^-, \delta^+ \phi^0_R, \delta^+ \phi^0_I, \phi^+ \delta^0_R, \phi^+ \delta^0_I \right)$
\[
\mathcal{M}_1 = \left( \begin{array}{cc} A^\prime & B^\prime \\ B^{\prime \dagger} & C^\prime \end{array} \right),
\]
with
\[
A^\prime = \left( \begin{array}{cccccc} 
2\lambda_2 + \frac{9}{2} \tilde\lambda_2 & 4\sqrt{3} \tilde\lambda_2 & \frac{3}{\sqrt{2}} \tilde\lambda_2 & -i\frac{3}{\sqrt{2}} \tilde\lambda_2 & \frac{\sqrt{3}}{2\sqrt{2}} \lambda_4 & -i\frac{\sqrt{3}}{2\sqrt{2}} \lambda_4 \\
4\sqrt{3} \tilde\lambda_2 & 2\lambda_2 + \frac{7}{2} \tilde\lambda_2 & 2\sqrt{6} \tilde\lambda_2 & -i2\sqrt{6} \tilde\lambda_2 & \frac{1}{\sqrt{2}} \lambda_4 & -i\frac{1}{\sqrt{2}} \lambda_4 \\
\frac{3}{\sqrt{2}} \tilde\lambda_2 & 2\sqrt{6} \tilde\lambda_2 & 2\lambda_2 + \frac{9}{2} \tilde\lambda_2 & 0 & \frac{\sqrt{3}}{4}\lambda_4 & -i\frac{\sqrt{3}}{4}\lambda_4 \\
i\frac{3}{\sqrt{2}} \tilde\lambda_2 & i2\sqrt{6} \tilde\lambda_2 & 0 & 2\lambda_2 + \frac{9}{2} \tilde\lambda_2 & i\frac{\sqrt{3}}{4}\lambda_4 & \frac{\sqrt{3}}{4}\lambda_4 \\
\frac{\sqrt{3}}{2\sqrt{2}} \lambda_4 & \frac{1}{\sqrt{2}} \lambda_4 & \frac{\sqrt{3}}{4}\lambda_4 & -i\frac{\sqrt{3}}{4}\lambda_4 & 2\lambda_1 & 0 \\
i\frac{\sqrt{3}}{2\sqrt{2}} \lambda_4 & i\frac{1}{\sqrt{2}} \lambda_4 & i\frac{\sqrt{3}}{4}\lambda_4 & \frac{\sqrt{3}}{4}\lambda_4 & 0 & 2\lambda_1
\end{array} \right)
\]

\[
B^\prime = \begin{pmatrix}
0_{4 \times 5} \\
\begin{array}{cc}
\begin{matrix}
\sqrt{6}\lambda_5 & \sqrt{3}\lambda_5 & -i\sqrt{3}\lambda_5 \\
-i\sqrt{6}\lambda_5 & -i\sqrt{3}\lambda_5 & -\sqrt{3}\lambda_5
\end{matrix}
& 0_{2 \times 2}
\end{array}
\end{pmatrix}
\]

\[
C^\prime = \left( \begin{array}{ccccc} 
\lambda_3 + \frac{1}{4} \lambda_4 & \frac{1}{\sqrt{2}} \lambda_4 & -i\frac{1}{\sqrt{2}} \lambda_4 & 0 & 0 \\
\frac{1}{\sqrt{2}} \lambda_4 & \lambda_3 + \frac{1}{4} \lambda_4 & 0 & \frac{\sqrt{3}}{4} \lambda_4 & i\frac{\sqrt{3}}{4} \lambda_4 \\
i\frac{1}{\sqrt{2}} \lambda_4 & 0 & \lambda_3 + \frac{1}{4} \lambda_4 & -i\frac{\sqrt{3}}{4} \lambda_4 & \frac{\sqrt{3}}{4} \lambda_4 \\
0 & \frac{\sqrt{3}}{4} \lambda_4 & i\frac{\sqrt{3}}{4} \lambda_4 & \lambda_3 - \frac{3}{4} \lambda_4 & 0 \\
0 & -i\frac{\sqrt{3}}{4} \lambda_4 & \frac{\sqrt{3}}{4} \lambda_4 & 0 & \lambda_3 - \frac{3}{4} \lambda_4
\end{array} \right)
\]

\item \underline{$q=0$:} \footnotesize $\left(\delta^{3+}\delta^{3-}, \delta^{2+}\delta^{2-}, \delta^+ \delta^-, \delta^+ \phi^-, \phi^+ \delta^-, \phi^+ \phi^-, \frac{\delta^0_R \delta^0_R}{\sqrt{2}}, \delta^0_R \delta^0_I , \frac{\delta^0_I \delta^0_I}{\sqrt{2}} ,\delta^0_R \phi^0_R, \delta^0_R \phi^0_I, \delta^0_I \phi^0_R, \delta^0_I \phi^0_I, \frac{\phi^0_R \phi^0_R}{\sqrt{2}}, \phi^0_R \phi^0_I, \frac{\phi^0_I \phi^0_I}{\sqrt{2}} \right)$
\small

\[
\mathcal{M}_0 = \left( \begin{array}{cc} A & B \\ C & D \end{array} \right),
\]
with
\[
A = \left( \begin{array}{ccccccc}
4\lambda_2+9\tilde\lambda_2 & 2\lambda_2+\frac{9}{2}\tilde\lambda_2 & 2\lambda_2-\frac{3}{2}\tilde\lambda_2 & 0 & 0 & \lambda_3 + \frac{3\lambda_4}{4} & \frac{1}{\sqrt{2}}(2\lambda_2 - \frac{9}{2}\tilde{\lambda}_2) \\
2\lambda_2+\frac{9}{2}\tilde\lambda_2 & 4\lambda_2+\tilde\lambda_2 & 2\lambda_2+\frac{7}{2}\tilde\lambda_2 & 0 & 0 & \lambda_3 + \frac{1}{4}\lambda_4 & \frac{1}{\sqrt{2}}(2\lambda_2 - \frac{3}{2}\tilde{\lambda}_2) \\
2\lambda_2-\frac{3}{2}\tilde\lambda_2 & 2\lambda_2+\frac{7}{2}\tilde\lambda_2 & 4\lambda_2+\tilde\lambda_2 & 0 & 0 & \lambda_3- \frac{1}{4}\lambda_4 & \frac{1}{\sqrt{2}}(2\lambda_2 + \frac{9}{2}\tilde{\lambda}_2) \\
0 & 0 & 0 & \lambda_3 - \frac{\lambda_4}{4} & 0 & 0 & 0 \\
0 & 0 & 0 & 0 & \lambda_3 - \frac{\lambda_4}{4} & 0 & 0 \\
\lambda_3 + \frac{3}{4}\lambda_4 & \lambda_3 +\frac{1}{4}\lambda_4 & \lambda_3 - \frac{1}{4}\lambda_4 & 0 & 0 & 4\lambda_1 & \frac{1}{\sqrt{2}}(\lambda_3 - \frac{3}{4}\lambda_4) \\
\frac{1}{\sqrt{2}}(2\lambda_2 - \frac{9}{2}\tilde{\lambda}_2) & \frac{1}{\sqrt{2}}(2\lambda_2 - \frac{3}{2}\tilde{\lambda}_2) & \frac{1}{\sqrt{2}}(2\lambda_2 + \frac{9}{2}\tilde{\lambda}_2) & 0 & 0 & \frac{1}{\sqrt{2}}(\lambda_3 - \frac{3}{4}\lambda_4) & 3(\lambda_2 + \frac{9}{4}\tilde{\lambda}_2) \\
\end{array} \right)
\]

\[
B = \left( \begin{array}{cccccccccccccccc}
0 & \frac{1}{\sqrt{2}}(2\lambda_2-\frac{9\tilde{\lambda}_2}{2}) & 0 & 0 & 0 & 0 & \frac{1}{\sqrt{2}}(\lambda_3 - \frac{3}{4}\lambda_4) & 0 & \frac{1}{\sqrt{2}}(\lambda_3 - \frac{3}{4}\lambda_4) \\
0 & \frac{1}{\sqrt{2}}(2\lambda_2-\frac{3\tilde{\lambda}_2}{2}) & 0 & 0 & 0 & 0 & \frac{1}{\sqrt{2}}(\lambda_3 - \frac{1}{4}\lambda_4) & 0&\frac{1}{\sqrt{2}}(\lambda_3 - \frac{1}{4}\lambda_4) \\
0 & \frac{1}{\sqrt{2}}(2\lambda_2 + \frac{9}{2}\tilde{\lambda}_2) & 0 & 0 & 0 & 0 & \frac{1}{\sqrt{2}}(\lambda_3 + \frac{1}{4}\lambda_4) & 0 & \frac{1}{\sqrt{2}}(\lambda_3 + \frac{1}{4}\lambda_4) \\
0 & 0 & \frac{\sqrt{3}}{4}\lambda_4 & -i\frac{\sqrt{3}}{4} \lambda_4 & i\frac{\sqrt{3}}{4} \lambda_4 & \frac{\sqrt{3}}{4}\lambda_4 & \frac{\sqrt{3}}{\sqrt{2}}\lambda_5 & i \sqrt{3}\lambda_5 & -\frac{\sqrt{3}}{\sqrt{2}}\lambda_5 \\
0 & 0 & \frac{\sqrt{3}\lambda_4}{4} & i\frac{\sqrt{3}\lambda_4}{4}  & -i\frac{\sqrt{3}\lambda_4}{4}  & \frac{\sqrt{3}}{4}\lambda_4 & \frac{\sqrt{3}}{\sqrt{2}}\lambda_5 & -i \sqrt{3}\lambda_5 & -\frac{\sqrt{3}}{\sqrt{2}}\lambda_5 \\
0 & \frac{1}{\sqrt{2}}(\lambda_3 - \frac{3}{4}\lambda_4) & 0 & 0 & 0 & 0 & \sqrt{2}\lambda_1 & 0 &  \sqrt{2}\lambda_1\\
0 & \lambda_2 + \frac{9}{4}\tilde{\lambda}_2 & 0 & 0 & 0 & 0 & \frac{1}{2}(\lambda_3 + \frac{3}{4}\lambda_4) & 0 & \frac{1}{2}(\lambda_3 + \frac{3}{4}\lambda_4) \\
\end{array} \right)
\]

\[
C = \left( \begin{array}{ccccccc}
0& 0 &0 & 0& 0 & 0 & 0 \\
\frac{1}{\sqrt{2}}(2\lambda_2-\frac{9\tilde{\lambda}_2}{2}) & \frac{1}{\sqrt{2}}(2\lambda_2-\frac{3\tilde{\lambda}_2}{2}) & \frac{1}{\sqrt{2}}(2\lambda_2 + \frac{9}{2}\tilde{\lambda}_2) & 0 & 0 & \frac{1}{\sqrt{2}}(\lambda_3 - \frac{3}{4}\lambda_4) & \lambda_2 + \frac{9}{4}\tilde{\lambda}_2 \\
0& 0 &0 & \frac{\sqrt{3}}{4}\lambda_4& \frac{\sqrt{3}}{4}\lambda_4 & 0 & 0 \\
0& 0 &0 & i\frac{\sqrt{3}}{4} \lambda_4& -i\frac{\sqrt{3}}{4} \lambda_4 & 0 & 0 \\
0& 0 &0 & -i\frac{\sqrt{3}}{4} \lambda_4& i\frac{\sqrt{3}}{4} \lambda_4& 0 & 0 \\		
0& 0 &0 & \frac{\sqrt{3}}{4}\lambda_4& \frac{\sqrt{3}}{4}\lambda_4 & 0 & 0 \\	 
\frac{1}{\sqrt{2}}(\lambda_3 - \frac{3}{4}\lambda_4)& \frac{1}{\sqrt{2}}(\lambda_3 - \frac{1}{4}\lambda_4) &\frac{1}{\sqrt{2}}(\lambda_3 + \frac{1}{4}\lambda_4) & \frac{\sqrt{3}}{\sqrt{2}}\lambda_5& \frac{\sqrt{3}}{\sqrt{2}} \lambda_5&  \sqrt{2}\lambda_1& \frac{1}{2}(\lambda_3 + \frac{3}{4}\lambda_4) \\		 
0& 0 &0 & -i \sqrt{3}\lambda_5& i \sqrt{3}\lambda_5 & 0 & 0 \\		
\frac{1}{\sqrt{2}}(\lambda_3 - \frac{3}{4}\lambda_4) &\frac{1}{\sqrt{2}}(\lambda_3 - \frac{1}{4}\lambda_4)&\frac{1}{\sqrt{2}}(\lambda_3 + \frac{1}{4}\lambda_4) & -\frac{\sqrt{3}}{\sqrt{2}}\lambda_5& -\frac{\sqrt{3}}{\sqrt{2}}\lambda_5 & \sqrt{2}\lambda_1 & \frac{1}{2}(\lambda_3 + \frac{3}{4}\lambda_4)
\end{array} \right)
\]

\[
D = \left( \begin{array}{ccccccccc}
2\lambda_2+\frac{9\tilde{\lambda}_2}{2}  & 0 & 0 & 0 & 0 & 0 & 0 & 0 & 0 \\
0 & 3(\lambda_2 + \frac{9}{4}\tilde{\lambda}_2) & 0 & 0 & 0 & 0 & \frac{1}{2}( \lambda_3+\frac{3}{4}\lambda_4) & 0 & \frac{1}{2}(\lambda_3 + \frac{3}{4}\lambda_4) \\
0 & 0 & \lambda_3 + \frac{3\lambda_4}{4} & 0 & 0 & 0 & \frac{3}{\sqrt{2}}\lambda_5 & 0 & -\frac{3}{\sqrt{2}}\lambda_5 \\
0 & 0 & 0 & \lambda_3 + \frac{3\lambda_4}{4} & 0 & 0 & 0 & -3\lambda_5 & 0 \\
0 & 0 & 0 & 0 & \lambda_3+\frac{3}{4}\lambda_4 & 0 & 0 & 3\lambda_5  & 0 \\		
0 & 0 & 0 & 0 & 0 & \lambda_3+\frac{3}{4}\lambda_4 & \frac{3}{\sqrt{2}}\lambda_5  & 0 & -\frac{3}{\sqrt{2}}\lambda_5 \\	 
0 & \frac{1}{2}( \lambda_3+\frac{3}{4}\lambda_4) & \frac{3}{\sqrt{2}}\lambda_5 & 0 &0 & \frac{3}{\sqrt{2}}\lambda_5 & 3 \lambda_1 & 0 & \lambda_1 \\		 
0 & 0 & 0 & -3\lambda_5 & 3\lambda_5 & 0 & 0 & 2\lambda_1  & 0 \\		
0 & \frac{1}{2}(\lambda_3 + \frac{3}{4}\lambda_4) & -\frac{3}{\sqrt{2}}\lambda_5 & 0 & 0 & -\frac{3}{\sqrt{2}}\lambda_5 & \lambda_1 & 0 & 3\lambda_1		 
\end{array} \right)
\]

\end{itemize}

%% file: sections/appB.tex
\section{Minimization conditions}
\label{app:min}
Here we collect the minimization conditions corresponding to the stationary configurations discussed in Section~\ref{sec:globality}.

\begin{align*}
{\rm N1}: &~ \mu_\Phi^2 = \lambda_1 v_\Phi^2 + \frac{1}{8} \left(4\lambda_3 + 3\lambda_4 \right) v_\Delta^2 + \frac{3}{2} \lambda_5 v_\Phi v_\Delta, 
\quad
\mu_\Delta^2 = -(\lambda_2 +\frac{9}{4} \tilde{\lambda}_2) v_\Delta^2 - \frac{1}{8} \left( 4\lambda_3 + 3\lambda_4 \right) v_\Phi^2 - \frac{\lambda_5}{2}  \frac{v_\Phi^3}{v_\Delta}
\\[4pt]
{\rm N2}: &~ (\text{exists only for }\lambda_5 = 0)~ \mu_\Phi^2 = \lambda_1 v_\Phi^{\prime 2}
\\[4pt]
{\rm CB1}: &~ \mu_\Phi^2 = \lambda_1 v^2 + \frac{1}{8} \left( 4\lambda_3 + \lambda_4 \right) v_1^2,
\quad
\mu_\Delta^2 = -\left(\lambda_2 + \frac{1}{4}\tilde\lambda_2 \right) v_1^2 - \frac{1}{8} \left(4\lambda_3 + \lambda_4 \right) v^2
\\[4pt]
{\rm CB2}: &~ \mu_\Phi^2 = \lambda_1 v^2 + \frac{1}{8} \left( 4\lambda_3 - \lambda_4 \right) v_2^2,
\quad
\mu_\Delta^2 = -\left(\lambda_2 + \frac{1}{4} \tilde\lambda_2 \right) v_2^2 - \frac{1}{8} \left(4\lambda_3 - \lambda_4 \right) v^2
\\[4pt]
{\rm CB3}: &~ \mu_\Phi^2 = \lambda_1 v^2 + \frac{1}{8} \left( 4\lambda_3 - 3\lambda_4 \right) v_3^2,
\quad
\mu_\Delta^2 = -\left(\lambda_2 + \frac{9}{4} \tilde\lambda_2 \right) v_3^2 - \frac{1}{8} \left( 4\lambda_3 - 3\lambda_4 \right) v^2
\\[4pt] 
{\rm CB4}: &~ \mu_\Phi^2 = \lambda_1 v^2 + \frac{1}{8} \left( 4\lambda_3 + \lambda_4 \right) v_1^2 + \frac{1}{8} \left( 4\lambda_3 - \lambda_4 \right) v_2^2
\\
&\mu_\Delta^2 = -\lambda_2 (v_1^2 + v_2^2) - \frac{v^2}{6(v_1^2-v_2^2)} \left\{ \left(3\lambda_3 + \lambda_4\right) v_1^2 - \frac{1}{6} \left(3\lambda_3 - \lambda_4\right) v_2^2 \right\},
\quad
6\tilde\lambda_2 (v_1^2 - v_2^2)= \lambda_4 v^2
\\[4pt]
{\rm CB5}: &~ \mu_\Phi^2 = \lambda_1 v^2 + \frac{1}{8} \left(4\lambda_3 + \lambda_4\right) v_1^2 + \frac{1}{8} \left(4\lambda_3 - 3\lambda_4\right) v_3^2,
\quad
\mu_\Delta^2 = -\frac{1}{2}\lambda_3 v^2 - \lambda_2 \left( v_1^2 + v_3^2 \right),
\quad
2\tilde\lambda_2 \left( v_1^2 - 3v_3^2 \right) + \lambda_4 v^2 = 0
\\[4pt]
{\rm CB6}: &~ \mu_\Phi^2 = \lambda_1 v^2 + \frac{1}{2}\lambda_3 \left( v_2^2 + v_3^2 \right) - \frac{1}{8}\lambda_4 \left( v_2^2 + 3 v_3^2 \right),
\quad
\mu_\Delta^2 = -\lambda_2 \left( v_2^2 + v_3^2 \right) - \frac{1}{32} \left(16\lambda_3 - 3\lambda_4 \right) v^2 -\frac{9v^2 v_3^2}{32v_2^2} \lambda_4
\\
&~ 8\tilde\lambda_2 v_2^2 = \lambda_4 v^2
\\[4pt]
{\rm CB7}: &~ \mu_\Phi^2 = \lambda_1 v^2 + \frac{1}{2}\lambda_3 \left( v_1^2 + v_2^2 + v_3^2 \right),
\quad
\mu_\Delta^2 = -\lambda_2 \left( v_1^2 + v_2^2 + v_3^2 \right) - \frac{\lambda_3}{2} v^2,
\quad
\tilde\lambda_2 = 0, \quad \lambda_4 = 0
\\[4pt]
{\rm CB8}: &~ \mu_\Phi^2 = \lambda_1 v^2 + \frac{\lambda_3}{2} \left( v_0^2 + v_1^2 \right) + \frac{\lambda_4}{8} \left(3v_0^2 + v_1^2\right) + \frac{3}{2} \lambda_5 v v_0
\\
&~ \mu_\Delta^2 = -\lambda_2 \left(v_0^2 + v_1^2 \right) - \frac{1}{32} \left(16\lambda_3 + 3\lambda_4 \right) v^2  + \frac{9v^2v_0^2}{32v_1^2} \lambda_4 + \frac{v^3\left(9v_0^2 + v_1^2\right)}{16v_0v_1^2} \lambda_5,
\quad
8\tilde\lambda_2 v_0 v_1^2 + \lambda_4 v^2 v_0 + 2\lambda_5 v^3 = 0
\end{align*}
\begin{align*}
{\rm CB9}: &~ \mu_\Phi^2 = \lambda_1 v^2 + \frac{\lambda_3}{2} \left( v_0^2 + v_2^2 \right) + \frac{\lambda_4}{8} \left(3v_0^2 - v_2^2\right) + \frac{3}{2} \lambda_5 v v_0,
\quad
\mu_\Delta^2 = -\lambda_2 \left(v_0^2 + v_2^2 \right) - \frac{\lambda_3}{2} v^2 - \frac{v^3}{8v_0} \lambda_5
\\
&~ 2\tilde\lambda_2 v_0 \left( 3v_0^2 - v_2^2 \right) + \lambda_4 v^2 v_0 + \lambda_5 v^3 = 0
\\[4pt]
{\rm CB10}: &~ \mu_\Phi^2 =\lambda_1 v^2 + \frac{\lambda_3}{2} \left( v_0^2 + v_3^2 \right) + \frac{3\lambda_4}{8} \left(v_0^2 - v_3^2\right) + \frac{3}{2} \lambda_5 v v_0,
\quad
\mu_\Delta^2 = -\lambda_2 \left(v_0^2 + v_3^2 \right) - \frac{\lambda_3}{2} v^2 - \frac{v^3}{4v_0} \lambda_5
\\
&~ 18\tilde\lambda_2 v_0 \left(v_0^2 - v_3^2 \right) + 3\lambda_4 v^2 v_0 + 2\lambda_5 v^3 = 0
\\
{\rm CB11}: &~ \mu_\Phi^2 =\lambda_1 v^2 + \frac{\lambda_3}{2} \left( v_0^2 + v_1^2 + v_2^2 \right) + \frac{\lambda_4}{8} \left(-13 v_0^2 + v_1^2 - v_2^2 + \frac{2v_0 \left(30 v_0^3 v_2 - 3 v_0 v_1^2 v_2 + 6\sqrt{3} v_1^2 v_2^2 - 2\sqrt{3}v_0^2 (5v_1^2 - 4v_2^2 )\right)}{6 v_0^2 v_2 - 3 v_2 \left( v_1^2 - v_2^2 \right) - 2 \sqrt{3} v_0 (v_1^2 - 2 v_2^2)}\right) 
\\
&~ \mu_\Delta^2 = -\lambda_2 \left(v_0^2 + v_1^2 + v_2^2 \right) - \frac{\lambda_3}{2} v^2 + \frac{\lambda_4}{8} \frac{3v_0^2 v_2 + 4 v_2 \left( v_1^2 + v_2^2 \right) + 2\sqrt{3} v_0 (v_1^2 + 2 v_2^2)}{6 v_0^2 v_2 - 3 v_2 \left( v_1^2 - v_2^2 \right) - 2 \sqrt{3} v_0 (v_1^2 - 2 v_2^2)} v^2
\\
&~ 2\tilde\lambda_2 \left( 6 v_0^2 v_2 - 3 v_2 \left( v_1^2 - v_2^2 \right) - 2 \sqrt{3} v_0 (v_1^2 - 2 v_2^2) \right)  + \lambda_4 v^2 v_2  = 0
\\
&~ 2\lambda_5 \left( 6 v_0^2 v_2 - 3 v_2 \left( v_1^2 - v_2^2 \right) - 2 \sqrt{3} v_0 (v_1^2 - 2 v_2^2) \right) v = \lambda_4 \left(- 6 v_0^3 v_2 + 2\sqrt{3} v_1^2 v_2^2 + 2\sqrt{3} v_0^2 (v_1^2 -4 v_2^2) + v_0 (7 v_1^2 v_2 - 8 v_2^3)\right)
\\[4pt]
{\rm CB12}: &~ \mu_\Phi^2 =\lambda_1 v^2 + \frac{\lambda_3}{2} \left( v_0^2 + v_1^2 + v_3^2 \right) + \frac{\lambda_4}{8} \frac{-27 v_0^4 + v_0^2 (51 v_1^2 - 45 v_3^2) + 2 (v_1^2 - 3 v_3^2)^2}{9 v_0^2 + 2v_1^2 - 6 v_3^2}
\\
&~ \mu_\Delta^2 = -\lambda_2 \left(v_0^2 + v_1^2 + v_3^2 \right) - \frac{\lambda_3}{2} v^2 +\frac{\lambda_4}{8} \frac{9 v_0^2 }{9 v_0^2 + 2v_1^2 - 6 v_3^2} v^2
\\
&~ \tilde\lambda_2 (9 v_0^2 + 2v_1^2 - 6 v_3^2) + \lambda_4 v^2  = 0,
\quad
2 \lambda_5 v (9 v_0^2 + 2v_1^2 - 6 v_3^2) + \lambda_4 (9 v_0^3 - 6 v_0 v_1^2 )  = 0
\\[4pt]
{\rm CB13}: &~ \mu_\Phi^2 =\lambda_1 v^2 + \frac{\lambda_3}{2} \left( v_0^2 + v_2^2 + v_3^2 \right) + \lambda_4 \frac{9v_0^4 + 21v_0^2 \left(v_2^2 -3v_3^2 \right) + 4 \left(v_2^4 + 3v_2^2v_3^2 \right)}{3v_0^2 - 4v_2^2}
\\
&~ \mu_\Delta^2 = -\lambda_2 \left(v_0^2 + v_2^2 + v_3^2 \right) - \frac{\lambda_3}{2} v^2 - \frac{3\lambda_4}{8} \frac{v_2^2 - 3v_3^2}{3v_0^2 - 4v_2^2} v^2
\\
&~ 2\tilde\lambda_2 \left( 3v_0^2 - 4v_2^2 \right) + \lambda_4 v^2 = 0, 
\quad
2\lambda_5 v \left( 3v_0^2 - 4v_2^2 \right) - 3\lambda_4v_0 \left( 2v_2^2 - 3v_3^2 \right) = 0
\\[4pt]
{\rm CB14}: &~ \mu_\Phi^2 =\lambda_1 v^2 + \frac{\lambda_3}{2} \left( v_0^2 + v_1^2 +  v_2^2 + v_3^2 \right),
\quad
\mu_\Delta^2 = -\lambda_2 \left(v_0^2 + v_1^2 + v_2^2 + v_3^2 \right) - \frac{\lambda_3}{2} v^2, \quad \tilde\lambda_2 = 0, \quad \lambda_4 = 0,  \quad \lambda_5 = 0.
\end{align*}

%% file: sections/appC.tex
\section{Potential differences between stationary configurations}
\label{app:potdiff}
Here we provide the explicit expressions for the potential differences between coexisting stationary configurations, which are used in the analysis of vacuum stability in Section~\ref{sec:globality}.
\begin{align*}
V_{CB1} - V_{N2} =&~ \frac{v_1^2}{32} \left\{ 8\mu_\Delta^2 + \left( 4\lambda_3 + \lambda_4 \right) v_\Phi^2 \right\} = \frac{v_1^2}{4} \left(2m_{H^{\pm\pm}}^2 - m_{H^{\pm\pm\pm}}^2 \right)
\\
V_{CB2} - V_{N2} =&~ \frac{v_2^2}{32} \left\{ 8\mu_\Delta^2 + \left( 4\lambda_3 - \lambda_4 \right) v_\Phi^2 \right\} = \frac{v_2^2}{4} m_{H^{\pm\pm}}^2
\\
V_{CB3} - V_{N2} =&~ \frac{v_3^2}{32} \left\{ 8\mu_\Delta^2 + \left( 4\lambda_3 - 3\lambda_4 \right) v_\Phi^2 \right\} = \frac{v_3^2}{4} m_{H^{\pm\pm\pm}}^2
\\
V_{CB4} - V_{N2} =&~ \frac{v_\Phi^2}{32} \left\{ 4\lambda_3(v_1^2 + v_2^2) + \lambda_4(v_1^2 - v_2^2) \right\} + \frac{\mu_\Delta^2}{4}(v_1^2 + v_2^2) 
= ~ \frac{1}{4} \left\{ v_1^2 (2m_{H^{\pm\pm}}^2 - m_{H^{\pm\pm\pm}}^2) + v_2^2 m_{H^{\pm\pm}}^2 \right\}
\\
V_{CB5} - V_{N2} =&~ \frac{v_\Phi^2}{32} \left\{ 4\lambda_3(v_1^2 + v_3^2) + \lambda_4(v_1^2 - 3 v_3^2) \right\} + \frac{\mu_\Delta^2}{4}(v_1^2 + v_3^2) 
= ~ \frac{1}{4} \left\{ v_1^2 (2m_{H^{\pm\pm}}^2 - m_{H^{\pm\pm\pm}}^2) + v_3^2 m_{H^{\pm\pm\pm}}^2 \right\}
\\
V_{CB6} - V_{N2} =&~ \frac{v_\Phi^2}{32} \left\{ 4\lambda_3(v_2^2 + v_3^2) -\lambda_4(v_2^2 + 3 v_3^2) \right\} + \frac{\mu_\Delta^2}{4}(v_2^2 + v_3^2) = \frac{1}{4}(v_2^2 m_{H^{\pm\pm}}^2 + v_3^2 m_{H^{\pm\pm\pm}}^2 )
\\
V_{CB7} - V_{N2} =&~ \frac{v_\Phi^2}{32} \left\{ 4\lambda_3(v_1^2 + v_2^2 + v_3^2) + \lambda_4(v_1^2 - v_2^2 - 3 v_3^2)  \right\} + \frac{\mu_\Delta^2}{4}(v_1^2 + v_2^2 + v_3^2) 
\\
= &~ \frac{1}{4} \left\{ v_1^2 (2m_{H^{\pm\pm}}^2 - m_{H^{\pm\pm\pm}}^2) + v_2^2 m_{H^{\pm\pm}}^2 + v_3^2 m_{H^{\pm\pm\pm}}^2 \right\} 
\end{align*}
\begin{align*}
V_{CB8} - V_{N2} =&~ \frac{v_\Phi^2}{32} \left\{ 4\lambda_3(v_0^2 + v_1^2) + \lambda_4(3 v_0^2 +v_1^2) + 12\lambda_5 v v_0 \right\} + \frac{\mu_\Delta^2}{4}(v_0^2 + v_1^2 )
\\
= &~ \frac{1}{4} \left\{ v_0^2(3 m_{H^{\pm\pm}}^2 - 2 m_{H^{\pm\pm\pm}}^2)  + v_1^2 (2 m_{H^{\pm\pm}}^2 - m_{H^{\pm\pm\pm}}^2)\right\} + \frac{3}{8} \lambda_5 vv_0v_\Phi^2
\\
V_{CB9} - V_{N2} =&~ \frac{v_\Phi^2}{32} \left\{ 4\lambda_3(v_0^2 + v_2^2) + \lambda_4(3 v_0^2 -v_2^2) + 12\lambda_5 v v_0 \right\} + \frac{\mu_\Delta^2}{4}(v_0^2 + v_2^2 )
\\
= &~ \frac{1}{4} \left\{ v_0^2(3 m_{H^{\pm\pm}}^2 - 2 m_{H^{\pm\pm\pm}}^2)  + v_2^2 m_{H^{\pm\pm}}^2 \right\} + \frac{3}{8} \lambda_5 vv_0v_\Phi^2
\\
V_{CB10} - V_{N2} =&~ \frac{v_\Phi^2}{32} \left\{ 4\lambda_3(v_0^2 + v_3^2) + 3\lambda_4(v_0^2 -v_3^2) + 12\lambda_5 v v_0 \right\} + \frac{\mu_\Delta^2}{4}(v_0^2 + v_3^2 )
\\
= &~ \frac{1}{4} \left\{ v_0^2(3 m_{H^{\pm\pm}}^2 - 2 m_{H^{\pm\pm\pm}}^2)  + v_3^2 m_{H^{\pm\pm}}^2 \right\} + \frac{3}{8} \lambda_5 vv_0v_\Phi^2
\\
V_{CB11} - V_{N2} =&~ \frac{v_\Phi^2}{32} \left\{ 4\lambda_3(v_0^2 + v_1^2 + v_2^2) + \lambda_4(3v_0^2 + v_1^2 - v_2^2) + 12\lambda_5 v v_0 \right\} + \frac{\mu_\Delta^2}{4}(v_0^2 + v_1^2 + v_2^2)
\\
= &~ \frac{1}{4} \left\{ v_0^2(3 m_{H^{\pm\pm}}^2 - 2 m_{H^{\pm\pm\pm}}^2) + v_1^2(2 m_{H^{\pm\pm}}^2 - m_{H^{\pm\pm\pm}}^2) + v_2^2 m_{H^{\pm\pm}}^2 \right\} + \frac{3}{8} \lambda_5 vv_0v_\Phi^2
\\
V_{CB12} - V_{N2} =&~ \frac{v_\Phi^2}{32} \left\{ 4\lambda_3(v_0^2 + v_1^2 + v_3^2) + \lambda_4(3v_0^2 + v_1^2 - 3v_3^2) + 12\lambda_5 v v_0 \right\} + \frac{\mu_\Delta^2}{4}(v_0^2 + v_1^2 + v_3^2)
\\
= &~ \frac{1}{4} \left\{ v_0^2(3 m_{H^{\pm\pm}}^2 - 2 m_{H^{\pm\pm\pm}}^2) + v_1^2(2 m_{H^{\pm\pm}}^2 - m_{H^{\pm\pm\pm}}^2) + v_3^2 m_{H^{\pm\pm\pm}}^2 \right\} + \frac{3}{8} \lambda_5 vv_0v_\Phi^2
\\
V_{CB13} - V_{N2} =&~ \frac{v_\Phi^2}{32} \left\{ 4\lambda_3(v_0^2 + v_2^2 + v_3^2) + \lambda_4(3v_0^2 - v_2^2 - 3v_3^2) + 12\lambda_5 v v_0 \right\} + \frac{\mu_\Delta^2}{4}(v_0^2 + v_1^2 + v_3^2)
\\
= &~ \frac{1}{4} \left\{ v_0^2(3 m_{H^{\pm\pm}}^2 - 2 m_{H^{\pm\pm\pm}}^2) + v_2^2 m_{H^{\pm\pm}}^2 + v_3^2 m_{H^{\pm\pm\pm}}^2 \right\} + \frac{3}{8} \lambda_5 vv_0v_\Phi^2
\\
V_{CB14} - V_{N2} =&~ \frac{1}{8} \left( v_0^2 + v_1^2 + v_2^2 + v_3^2 \right) \left( \lambda_3 v_\phi^2 + 2\mu_\Delta^2 \right) = \frac{1}{8} \left( v_0^2 + v_1^2 + v_2^2 + v_3^2 \right) \left(3m_{H^{\pm\pm}}^2 - m_{H^{\pm\pm\pm}}^2 \right).
\end{align*}
Here, $m_{H^{\pm\pm}}$, $m_{H^{\pm\pm\pm}}$ denote the physical scalar masses evaluated at the stationary point $N2$. 

\begin{align*}
V_{CB1} - V_{N1} = & -\frac{\tilde\lambda_2}{2} v_1^2 v_\Delta^2 -\frac{\lambda_4}{16} \left( v^2 v_\Delta^2 + v_1^2 v_\Phi^2 \right) - \frac{\lambda_5}{8} \frac{v_\Phi}{v_\Delta} \left(3 v^2 v_\Delta^2 + v_1^2 v_\Phi^2 \right)
\\
V_{CB2} - V_{N1} = & -\frac{\tilde\lambda_2}{2} v_2^2 v_\Delta^2 -\frac{\lambda_4}{8} \left( v^2 v_\Delta^2 + v_2^2 v_\Phi^2 \right) - \frac{\lambda_5}{8} \frac{v_\Phi}{v_\Delta} \left(3 v^2 v_\Delta^2 + v_2^2 v_\Phi^2 \right)
\\
V_{CB3} - V_{N1} = &  -\frac{3\lambda_4}{16} \left( v^2 v_\Delta^2 + v_3^2 v_\Phi^2 \right) - \frac{\lambda_5}{8} \frac{v_\Phi}{v_\Delta} \left(3 v^2 v_\Delta^2 + v_3^2 v_\Phi^2 \right)
\\
V_{CB4} - V_{N1} = & - \frac{3 \tilde\lambda_2}{4} v_2^2 v_\Delta^2 + \lambda_4 \frac{v^2 v_\Delta^2 (-7 v_1^2 + 8 v_2^2) - 3 v_\Phi^2(v_1^4 + v_1^2 v_2^2 - 2 v_2^4)}{48 (v_1^2 - v_2^2)} - \frac{\lambda_5}{8} \frac{v_\Phi}{v_\Delta} \left\{3 v^2 v_\Delta^2 +(v_1^2 +v_2^2) v_\Phi^2 \right\}
\\
V_{CB5} - V_{N1} = & - \frac{9\tilde\lambda_2}{8} v_3^2 v_\Delta^2 + \frac{\lambda_4}{16} \left\{ \frac{3 v^2 v_1^2 v_\Delta^2 }{v_1^2 - 3v_3^2} - v_\Phi^2(v_1^2 + 3 v_3^2)\right\} - \frac{\lambda_5}{8} \frac{v_\Phi}{v_\Delta} \{3 v^2 v_\Delta^2  + (v_1^2 + v_3^2)v_\Phi^2\}
\\
V_{CB6} - V_{N1} = & - \frac{3\tilde\lambda_2}{8} (2 v_2^2 + 3 v_3^2) v_\Delta^2 + \frac{\lambda_4}{16} \left\{ 
\frac{3 v^2 v_\Delta^2}{4 v_2^2} (-2 v_2^2 + 3 v_3^2) - v_\Phi^2 (2 v_2^2 + 3 v_3^2)\right\} - \frac{\lambda_5}{8} \frac{v_\Phi}{v_\Delta} \left\{3 v^2 v_\Delta^2  + (v_2^2 + v_3^2)v_\Phi^2\right\} 
\\
V_{CB7} - V_{N1} = & - \frac{3\tilde\lambda_2}{8} v_\Delta^2 (2 v_2^2 + 3 v_3^2)- \frac{\lambda_4}{16} (v_2^2 + 2 v_2^2 + 3v_3^2) v_\Phi^2 - \frac{\lambda_5}{8} \frac{v_\Phi}{v_\Delta} \left\{3 v^2 v_\Delta^2  + (v_1^2 + v_2^2 + v_3^2)v_\Phi^2 \right\} 
\\
V_{CB8} - V_{N1} = & - \frac{\lambda_4}{16} v_1^2 v_\Phi^2 + \frac{\lambda_5}{8v_0v_\Delta} \left\{ v^3 v_\Delta^3 - 3 v^2 v_0 v_\Delta^2 v_\Phi + 3 v v_0^2 v_\Delta v_\Phi^2 - v_0(v_0^2 + v_1^2 ) v_\Phi^3 \right\}
\\
V_{CB9} - V_{N1} = & - \frac{3\tilde\lambda_2}{4} v_2^2 v_\Delta^2 - \frac{\lambda_4}{8} v_2^2 v_\Phi^2 + \frac{\lambda_5}{8v_0v_\Delta} \left\{ v^3 v_\Delta^3 - 3 v^2 v_0 v_\Delta^2 v_\Phi + 3 v v_0^2 v_\Delta v_\Phi^2 - v_0(v_0^2 + v_2^2 ) v_\Phi^3 \right\}
\\
V_{CB10} - V_{N1} = &- \frac{9\tilde\lambda_2}{8} v_3^2 v_\Delta^2 - \frac{3\lambda_4}{16} v_3^2 v_\Phi^2 + \frac{\lambda_5}{8v_0v_\Delta} \left\{ v^3 v_\Delta^3 - 3 v^2 v_0 v_\Delta^2 v_\Phi + 3 v v_0^2 v_\Delta v_\Phi^2 - v_0(v_0^2 + v_3^2 ) v_\Phi^3 \right\}
\\
V_{CB11} - V_{N1} = & - \frac{3\tilde\lambda_2}{4} v_2^2 v_\Delta^2 - \frac{\lambda_5}{8} \frac{v_\Phi}{v_\Delta} \left\{v^2 v_\Delta^2 - 2v v_0 v_\Delta v_\Phi + (v_0^2 + v_1^2 + v_2^2) v_\Phi^2\right\} + \frac{f_\Delta v_\Delta^2 + f_{\Delta\Phi} v_\Delta v_\Phi + f_\Phi v_\Phi^2}{16 \left\{6 v_0^2 v_2 - 3 v_1^2 v_2 + 3 v_2^3 - 2 \sqrt{3} v_0 (v_1^2 - 2 v_2^2)\right\}}
\end{align*}
\begin{align*}
V_{CB12} - V_{N1} = &  - \frac{9\tilde\lambda_2}{8} v_3^2 v_\Delta^2 - \frac{\lambda_5}{8} \frac{v_\Phi}{v_\Delta} \left\{v^2 v_\Delta^2 - 2v v_0 v_\Delta v_\Phi + (v_0^2 + v_1^2 + v_3^2) v_\Phi^2\right\}
\\
& -\frac{\lambda_4}{16} \frac{3vv_\Delta(3 v_0^2 - 2 v_1^2)(v v_\Delta - 2 v_0 v_\Phi) + \left\{(9 v_0^4 + 3 v_0^2 (v_1^2 + 9 v_3^2) + 2 (v_1^4 - 9 v_3^4)\right\} v_\Phi^2}{9 v_0^2 + 2 v_1^2 - 6 v_3^2}
\\
V_{CB13} - V_{N1} = & - \frac{3\tilde\lambda_2}{8} (2 v_2^2 + 3 v_3^2) v_\Delta^2 - \frac{\lambda_5}{8} \frac{v_\Phi}{v_\Delta} \left\{v^2 v_\Delta^2 - 2v v_0 v_\Delta v_\Phi + (v_0^2 + v_2^2 + v_3^2) v_\Phi^2\right\}
\\
& + \frac{\lambda_4}{16} \frac{3vv_\Delta(2 v_2^2 -3 v_3^2)(v v_\Delta - 2v_0 v_\Phi) + 2 (4 v_2^4 - 9v_0^2 v_3^2 + 6 v_2^2 v_3^2) v_\Phi^2}{3 v_0^2 - 4 v_2^2},
\\
V_{CB14} - V_{N1} = & - \frac{3\tilde\lambda_2}{8} (2 v_2^2 + 3 v_3^2) v_\Delta^2 -\frac{\lambda_4}{16} (v_1^2 + 2 v_2^2 + 3v_3^2)v_\Phi^2 - \frac{ \lambda_5}{8} \frac{v_\Phi}{v_\Delta} \left\{ v^2 v_\Delta^2 - 2 v v_0 v_\Delta v_\Phi + v_\Phi^2(v_0^2 + v_1^2 +  v_2^2 + v_3^2) \right\},
\end{align*}
where
\begin{align*}
f_\Delta =& v^2 \left\{-6 v_0^2 v_2 + 7 v_1^2 v_2 - 8 v_2^3 + 2 \sqrt{3} v_0 (v_1^2 - 4 v_2^2)\right\}
\\
f_{\Delta\Phi} =& - 2 v \left\{-6 v_0^3 v_2 + 2 \sqrt{3} v_1^2 v_2^2 + 2 \sqrt{3} v_0^2 (v_1^2 - 4 v_2^2) + v_0 (7 v_1^2 v_2 - 8 v_2^3)\right\}
\\
f_\Phi =& -6 v_0^4 v_2 + v_0^2 v_2 (v_1^2 - 20 v_2^2) + 2 \sqrt{3} v_0^3 (v_1^2 - 4 v_2^2) + 2 \sqrt{3} v_0 (v_1^4 + v_1^2 v_2^2 - 4 v_2^4) 
\\
& + 3 v_2 (v_1^4 + v_1^2 v_2^2 - 2 v_2^4).
\end{align*}